\documentclass{jfm}
\usepackage{graphicx}
\usepackage{epstopdf, epsfig}
\usepackage{amsmath}
\usepackage[table]{xcolor}
\usepackage{subcaption}
\usepackage{tabularx}
\usepackage{soul}
\usepackage{dblfloatfix}    
\usepackage{multicol}

\shorttitle{Wake Patterns of Oscillating Foils in Side-by-side Configuration}
\shortauthor{A. Gungor, M. S. U. Khalid and A. Hemmati}

\title{Classification of vortex patterns of oscillating foils in side-by-side configurations}

\author{Ahmet Gungor\aff{1}, Muhammad S. U. Khalid\aff{1}
	\and Arman Hemmati\aff{1} \corresp{\email{arman.hemmati@ualberta.ca}}}

\affiliation{\aff{1}Dept. of Mechanical Engineering, University of Alberta, Edmonton, AB \mbox{T6G 2R3}, Canada}

\begin{document}
\maketitle

\begin{abstract}
The unsteady hydrodynamics of two in-phase pitching foils arranged in side-by-side (parallel) configurations is examined for a range of Strouhal number and separation distance. Three distinct vortex patterns are identified in the Strohual number$-$separation distance phase map, which include separated wake, merged wake, and transitional-merged wake. Furthermore, a novel model is introduced based on fundamental flow variables including velocity, location, and circulation of dipole structures to quantitatively distinguish vortex patterns in the wake. The physical mechanism of wake merging process is also elucidated. When an oscillating foil experiences the jet deflection phenomenon, secondary structures shed from the primary street traverse in the other direction by making an angle with its parent vortex street. For parallel foils, secondary structures from the vortex street of the lower foil interact with the primary vortex street of the upper foil under certain kinematic conditions. This interaction triggers the wake merging process by influencing circulation of coherent structures in the upper part of the wake. It is unveiled that merging of the wakes leads to enhancements in propulsive efficiency by increasing thrust generation without a significant alteration in power requirements. These are attributed to the formation of a high-momentum jet by the merged vortex street, which possesses significantly larger circulation due to the amalgamation of the vortices, and major alterations in the evolution of leading edge vortices. Thus, flow physics that are thoroughly explored here are crucial in providing novel insights for future development of flow control techniques for efficient designs of bio-inspired underwater propulsors.

\end{abstract}


\section{Introduction}
\label{ch_introdcution}

Bio-mimicking is an innovative way of designing highly efficient robotic platforms for numerous engineering applications. Hence, understanding physical mechanisms employed by natural aquatic species is important to design next-generation autonomous swimming robots. An oscillating foil presents a generic model for the motion used by fish to propel \citep{anderson1998oscillating}. Various studies on such dynamic systems were conducted by researchers in the past few decades, addressing different aspects of the associated wake mechanics and hydrodynamic performance of oscillating foils. \cite{triantafyllou-optimal-thrust-1993} argued that Strouhal number, defined as $St={f}{A}/{U_{\infty}}$, was one of the governing parameter in fish-like swimming problems. Here, $f$ denotes the oscillation frequency, $A$ is the amplitude of oscillations, and $U_{\infty}$ shows the free-stream flow velocity. They showed that Strouhal number of highest propulsive efficiency of an oscillating foil overlapped with that of natural swimming for many fish, cetaceans, and marine mammals. 

At low Strouhal numbers, oscillating foils produce well-known B\'enard-von K\'arm\'an (BvK) vortex street. The wake transitions to reverse B\'enard-von K\'arm\'an vortex street with increasing $\mbox{St}$ \citep{koochesfahani-vortical-patterns-1989}. A further increase {in the value of $\mbox{St}$} triggers the symmetry breaking process in the wake, resulting in formation of deflected (asymmetric) B\'enard-von K\'arm\'an streets \citep{jones-knoller-betz-1998}. This phenomena has been extensively investigated in the literature for single foils \citep{liang-high-order-2011,ellenrieder-piv-asymmetric-2008, cleaver-bifurcating-flows-2012}. \cite{godoy-transitions-wake-2008} demonstrated that the flow parameters of flapping locomotion in nature coincides with the parameters of oscillating foils that produce deflected wakes. They further conjectured that natural swimmers and fliers could either utilize deflected wakes as their maneuvering strategy or avoid them during forward locomotion. \cite{godoy-model-symmetry-breaking-2009} argued that even though three-dimensional ($\mbox{3D}$) effects influence vortex dynamics of oscillating foils, the wake deflection was a quasi two-dimensional (Q2D) phenomenon. The wake deflection occurred when self-advection of the first shed dipole was strong enough to divert the main flow and subsequent dipoles away from the wake centerline. \cite{godoy-transitions-wake-2008} further proposed a model that quantitatively predicted wake deflection, considering offset between dipolar velocity and advection velocity of the dipoles. Although \mbox{$St$} has a significance on asymmetric characteristics of oscillating foils, amplitude of the oscillation is observed to considerably influence the attributions of deflected wakes. Symmetry breaking is triggered in the wake of oscillating foils at noticeably high oscillation amplitude for a fixed \mbox{$St$} \citep{godoy-transitions-wake-2008}. Further increase in the amplitude results in the transition from two-dimensional wake to three-dimensional wake, which suggests that transition from reverse BvK to deflected BvK is required for the formation of three-dimensional instabilities in the wake \citep{deng_dynamic_features_2015}. On the other hand, there are some conditions that inhibits the formation of deflected wakes. For example, interactions between the foil and shed vortices in the wake of an oscillating flexible foil may cancel the constitution of  asymmetric vortex street. Furthermore,  \cite{calderon-absence-asymmetric-wake-2014} showed that three-dimensional effects in the wake of finite span foils hinder the wake deflection, which is observed for the effectively infinite span foils under the same flow conditions. Three-dimensionality introduced by the tip vortex, which prevents the vortex coupling, and the symmetric circulation of interconnected vortex loops, which are due to the vortex topology of finite span foil, are two underlying reasons that were provided for cancellation of the deflection. 

Efficient propulsion through effective vorticity control implementation has been an outstanding challenge in engineering community for decades. Under inviscid and incompressible flow assumption, trapping a free vortex on the upper surface of a two-dimensional wing is theoretically capable of increasing lift generation through the introduction of a low pressure region \citep{huang_trapped_vortex_1982}. An adequately stabilized spanwise vortex can enhance the coefficient of lift up to 10 fold, which can be beneficial for design of short takeoff and landing (STOL) aircrafts \citep{rossow_trapped_vortex_1978}. \cite{saffman_trapped_vortex_1977} calculated the exact solution for potential flow over a flat plate with a free line vortex positioned on the upper boundary and estimated highly improved lift generation. Leading edge vortices (LEVs) substantially impact, often dominate the wake of simultaneously heaving and pitching foils, where their development could extensively amplify the propulsive performance of foils depending on their formation and shedding. These all are influenced by the foil kinematics. For instance, amalgamation of an LEV and a trailing edge vortex (TEV), which coincides with high efficiency and improved thrust generation, occurs when vortical structures are controlled using various parameters of foil kinematics, such as phase angle between heave and pitch, \mbox{$St$}, amplitude of heave motion or maximum angle of attack \citep{Anderson_phd_thesis_1996, anderson1998oscillating}. Likewise, three distinct vortex patterns are formed behind the foil simulatenosuly heaving and pitching in the wake of a D-section cylinder \citep{Gopalkrishnan_phd_thesis_1993, Gopalkrishnan_vorticity_control_1994, Shao_vorticity_control_2011}. Implementation of active vorticity control by dictating the flow kinematics yields constructive interaction mode, destructive interaction mode and expanding wake modes, which correspond to trough, peak and mixed responses in efficiency.

Natural swimmers and flyers are known to exploit physical mechanisms for their best interest to achieve most efficient way to displace themselves in a fluid medium. Reattachment of LEVs, formed by the flow separation due to dynamic stall, on the upper surface of the wing of hawkmoths or fruit flies during the downstroke of flapping greatly contributes to lift production \citep{Ellington_LEV_insect_1996, Birch_LEV_attachment_2001, Bomphrey_LEV_hawkmoth_2005}. Although balancing the body weight with enhanced lift production plays a crucial role in insect flight, it constitutes insignificant adversity for aquatic animals, owing to the presence of strong bouyant forces. The main concern for aquatic swimmers is to overcome the drag exerted by water, which is three orders of magnitude denser than air. \cite{borazjani_LEV_2013} carried out numerical simulations on self-propelled virtual swimmers with three different tail geometries inspired from mackerel body. They demonstrated that an attached LEV is formed on the body during locomotion settings that resemble natural swimming conditions for most fish. Evolution of the LEV is remarked to consequentially influence pressure distribution around the tail and the generated force for different tail shapes. In an experimental study, propulsive force of actively swimming bottlenose dolphins was calculated using digital particle image velocimetry (DPIV) measurements of vortex generated by the large amplitude fluke stroke of the dolphin \citep{fish_dolphing_DPIV_2014}. Effect of body shape (mackerel body or lamprey body) and swimming kinematics (anguilliform or carangiform) on the hydrodynamics of self-propelled virtual body/caudal fin swimmers was numerically examined by \cite{borazjani_swimming_kinematics_2010} for a range of Reynolds number. It is noted that the form and kinematics of swimmers differently impact the swimming efficiency in viscous, transitional and inertial regimes. \cite{liu_body-fin_2017} simulated a more complex model, which includes both fin-fin and body-fin interactions, by reconstructing body shape and kinematics of steady swimming crevalle jack using high-speed cameras. They demonstrated that posterior body vortices captured by the caudal fin strengthens LEVs around the fin, which produces most of the swimming thrust.

Fish schooling is defined as a behaviour seen in many fish species that appears as aggregation of a number of individuals and their collective navigation in the flow. Various reasons was propounded by evolutionary biologists and zoologists to explain this habit. These include but are not limited to improved success of finding a mate, effective defence strategy by confusing predators, and better chance of finding prospection. A fundamental question is raised in the minds of engineers interested in bio-inspiration: ``Could coordinated swimming enhance the propulsive performance of an individual swimmer?" One of the pioneer works, which addresses this question, was presented by \cite{weihs-hydromechanics-fish-schooling-1973}. They argued, based on a highly idealized, two-dimensional, and inviscid model, that individuals in schooling formation may enjoy hydrodynamic benefits if the spacing and synchronization between swimmers is adequately adjusted. \cite{daghooghi-rectangular-2015} conducted large-eddy simulations of self-propelled synchronized mackerels in a variety of infinite rectangular schooling patterns. They observed that schooling fish enjoy significant enhancements in swimming speed without more power requirements. They achieve this through exploitation of the channeling effect. \cite{ashraf-syncronization-2016, ashraf-simple-phalanx-2017} experimentally approached this problem by examining the swimming of two red nose tetra fish in a shallow-water tunnel with controlled velocity. By tracking the kinematics of fish using stereoscopic video recordings, it is demonstrated that the possibility of fish to locomote in side-by-side configuration and synchronize their tail beat frequencies is strongly correlated with the increasing speed of water. This suggests that in-phase or out-of-phase synchronization of the collectively swimming fish in parallel with a proper spacing provides intensified propulsive performance during demanding flow conditions.

Fish schools are often modeled using multiple oscillating rigid hydrofoils arranged in different configurations due to simplicity that it offers. \cite{boschitsch-in-line-tandem-2014} carried out experiments on the propulsive performance and wake structures of two pitching foils in an in-line configuration for a range of separation distances and phase differences. They observed that both the performance and wake structures of the front foil were affected by the presence of the downstream foil only for considerably small separation distances. They distinguished two different wake modes: branched and coherent. In the coherent mode, a single vortex street is formed behind the follower foil whose time-averaged wake corresponds to a single high-momentum jet. The branched mode, on the other hand, have two angled high-momentum jets in its time-averaged wake. The peaks in the thrust, power and efficiency coincide with the coherent mode wakes while the branched mode wakes are associated with the troughs. Recently, \cite{lagopoulos2020deflected} focused more on the wake deflection and production of side force by simultaneously heaving and pitching foils in an in-line configuration. They identified three distinct vortex patterns in the wake and showed that wake deflection introduced by the upstream foil could be eliminated due to the presence of the downstream body. 

\cite{dewey-propulsive-performance-2014} qualitatively examined vortex patterns of in-phase, mid-phase, and out-of-phase pitching foils in side-by-side configurations for a fixed separation distance and Strouhal number. They then proposed models of wake development for each case. To this end, it was shown that in-phase, out-of-phase, and mid-phase pitching foils produced merging symmetric, diverging symmetric, and asymmetric wakes, respectively. More recently, \cite{gungor-asymmetry} reported on numerical studies of in-phase and out-of-phase pitching foils in parallel arrangements at different Strouhal numbers that maintain a constant gap between them. They showed that the two foils produced quasi-steady symmetric wakes for both phase difference at low $\mbox{St}$, whereas asymmetric-to-symmetric and symmetric-to-asymmetric transitions were observed in the wake at high $\mbox{St}$ for in-phase and out-of-phase oscillations, respectively. A similar symmetry breaking phenomenon in the wake of foils, performing out-of-phase oscillations in a parallel configuration, was observed by \cite{bao-anti-phase-foils-2017} and \cite{zhang-locomotion-bioinspired-2018}. However, they demonstrated the asymmetric wake at a single time instant without examining the transient formation process of their asymmetry.

Although there are a few studies that demonstrate the development of vortex structures behind parallel foils, there are none that provide quantitative explanations for the vortex interactions in the wake. Furthermore, studies focused on explaining vortex patterns in the wake mostly overlooked unsteady interactions and their impact on propulsive performance, which are expected at high $\mbox{St}$. In this study, we examine merging of the vortex streets in the wake of in-phase pitching foils in side-by-side arrangement at a range of $\mbox{St}$ and separation distance inspired from fish schools. The Reynolds number of the flow ($Re=U_{\infty}c/\nu$, where, c is the chord length of the foil and $\nu$ is dynamic viscosity of the fluid) is fixed at $Re=4000$ considering wake patterns of oscillating foils reach a plateau after $Re\geq1000$ \citep{das-existence-sharp-transition-2016} and their propulsive performance exhibits negligible alteration after $Re \geq 4000$ \citep{senturk-reynolds-scaling-2019}. Therefore, this paper aims to illuminate three novel points that are currently missing in literature: (i) quantification and classification of vortex patterns behind two parallel foils undergoing in-phase pitching oscillations, (ii) physical mechanisms, governing the wake merging phenomenon, and (iii) influence of the merger on propulsive performance of the system. For this purpose, this paper is structured as follows. A description of dynamic system composed of two pitching foils and our numerical setup is provided in section \ref{ch_methodology}. Section \ref{ch_results} includes the results on the wakes of parallel foils and discussions concerning the vortex patterns and wake merging phenomena, which is followed by main conclusions in section \ref{cp_conclusion}.

\section{Methodology}
\label{ch_methodology}

\begin{figure}
	\centering
	\includegraphics[width=1\textwidth]{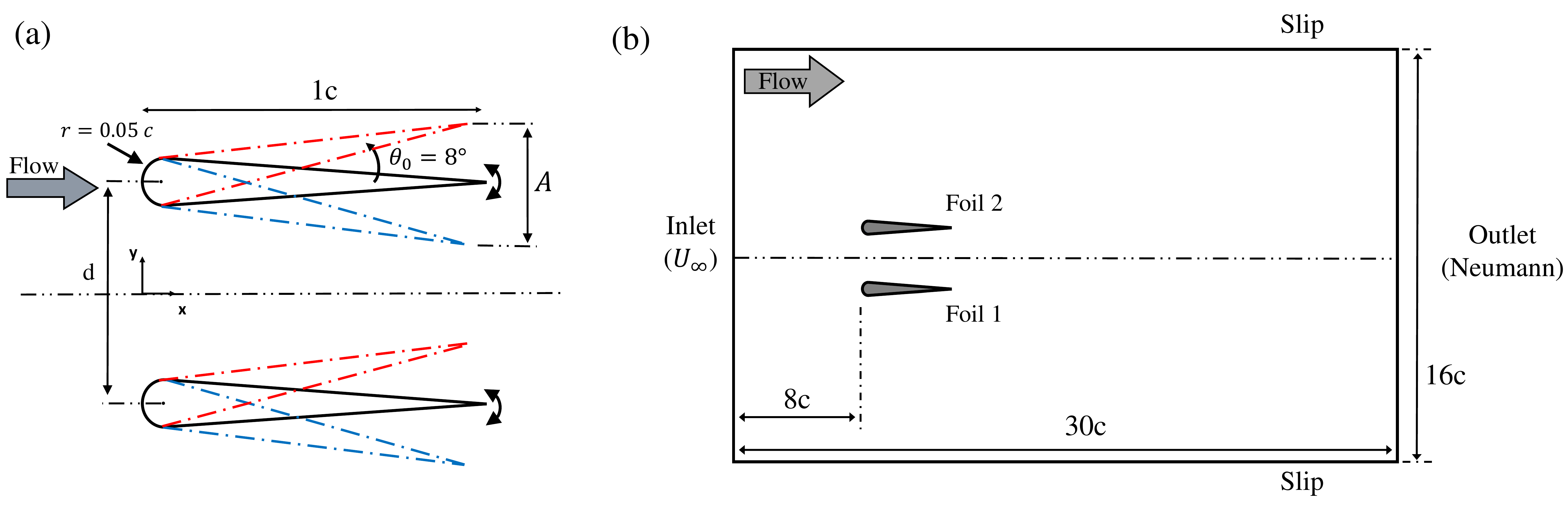}%
	\caption{Demonstration of (a) the pitching motion (b) two-dimensional computational domain with boundary conditions (not to scale).}
	\label{fig_setup}
\end{figure}

The flow around two oscillating, rigid teardrop foils in a side-by-side configuration is numerically simulated using OpenFOAM. For this purpose, Navier-Stokes equations are directly solved using pimpleDyMFOAM solver, which is an incompressible transient flow solver for systems requiring dynamic grids. The solver utilizes PIMPLE algorithm, which is a hybrid of PISO (Pressure-Implicit with Splitting Operators) and SIMPLE (Semi Implicit Method for Pressure Linked Equations). The time-step size is adequately selected to limit Courant number of the flow below 0.8 throughout the domain. It is achieved by using over 3500 time-steps per oscillation cycles. The divergence terms of Navier-Stokes equations are discretized using upwind-biased, second-order accurate ``Linear Upwind" technique. Second-order, implicit backward time method is employed for temporal terms. The convergence criterion, which is the residual of velocity components and pressure in the momentum equations, is set to $10^{-5}$.   

\begin{figure}
	\centering
	\includegraphics[width=1\textwidth]{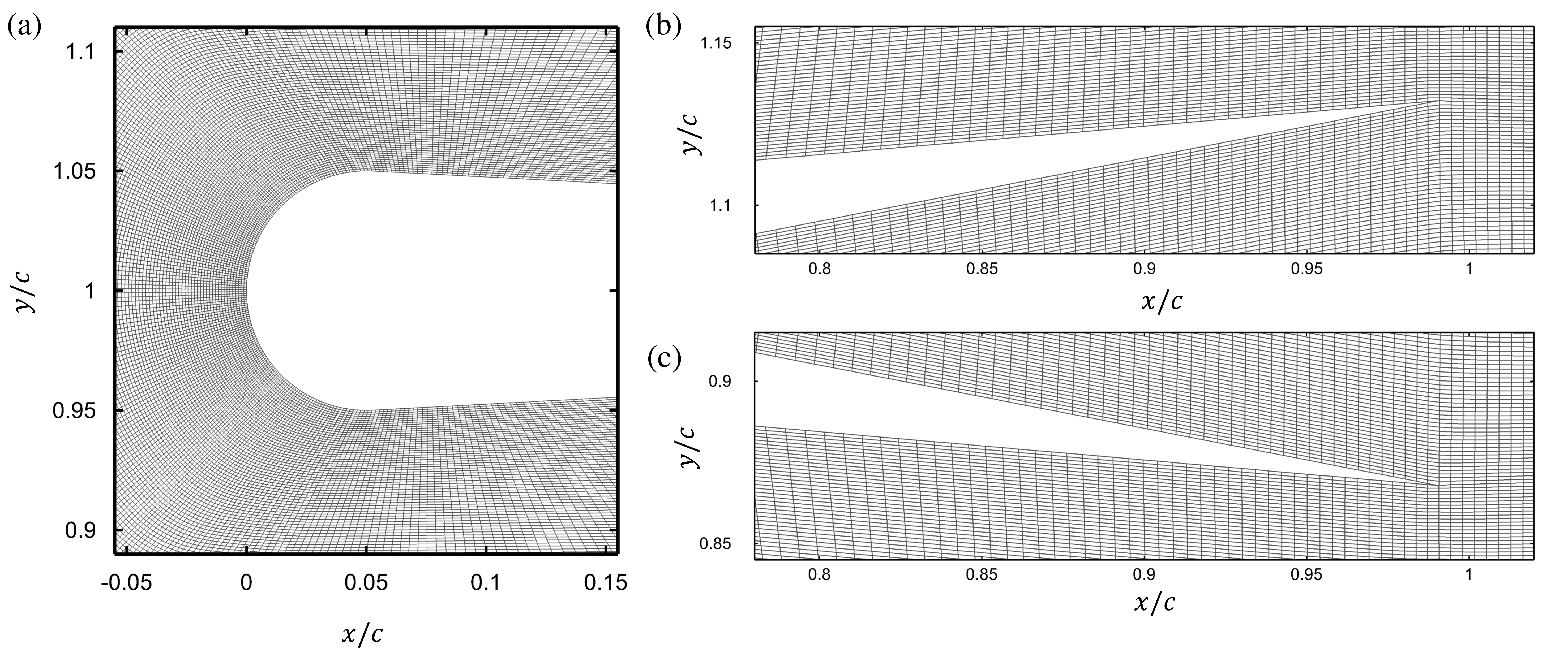}
	\caption{Details of the spatial grid around (a) leading edge (b) trailing edge during upstroke ($\theta=8^{\circ}$) (c) trailing edge during downstroke ($\theta=-8^{\circ}$).}
	\label{fig_grid}
\end{figure}

Both foils, Foil 1 (lower foil) and Foil 2 (upper foil), have chord-lengths of $c$ and semicircular leading edges with radii of $0.05c$. They perform pure pitching motion, which is mathematically defined as: 
\begin{equation}
\theta_1(t)=\theta_0\sin(2\pi ft),
\end{equation}
\begin{equation}
\theta_2(t)=\theta_0\sin(2\pi ft+\phi).
\end{equation}
Here, $\theta_0$ is the pitching amplitude, $t$ is time, and $\phi$ is the phase difference between the two foils. In-phase pitching condition are imposed by setting $\phi=0$, and $\theta_0$ is fixed at $8^\circ$. A schematic representation of the in-phase pitching is provided in Figure \ref{fig_setup}a. The grid is morphed by the solver in each timestep in order to ensure the pitching motion, while maintaining its quality. The separation distance between the foils ($d$) is varied from $0.5c$ to $2c$ with increments of $0.5c$ and is non-dimensionalized by $c$, i.e., $0.5<d^*=d/c<2$. Moreover, $\mbox{St}$ ranges from $0.15$ to $0.5$ for each value of $d^*$. This parameter space ($d^*$ extend) includes the range of separation distance between two red nose tetra fish, synchronously swimming in side-by-side arrangement, during the fish tank experiments by \cite{ashraf-syncronization-2016, ashraf-simple-phalanx-2017}. The $\mbox{St}$ space covers the formation of BvK, reverse BvK, and reverse BvK regimes in the wake of single oscillating foils \citep{godoy-transitions-wake-2008} and natural swimming $\mbox{St}$ of various fish species \citep{triantafyllou-optimal-thrust-1993}.

A computational domain, similar to the one reported in our previous work \citep{gungor-scaling, gungor-fish-swimming_wake} is employed in this study, which follows the experiments of \cite{dewey-propulsive-performance-2014}. It extends $30c$ in the streamwise (x$-$) direction and $16c$ in the cross-flow (y$-$) direction. Also, the leading edge of the foils are placed $8c$ away from the inlet. Neumann condition for both pressure and velocity are applied at the outlet boundary, while a uniform velocity ($u=U_{\infty}, v=0, w=0$) is  prescribed to the inlet boundary. Boundary conditions for the upper and lower walls, and foil surfaces are selected to be slip and no-slip, respectively.

A non-homogeneous spatial grid, consisting of $7.87\times10^5$ hexahedral elements, is generated to simulate the flow. The grid is most refined around the foils with $600$ nodes on the surface of each foil, which is consistent with the numerical setup of \cite{senturk-reynolds-scaling-2019}. The grid size expands towards the {boundaries} without exceeding the expansion ratio of $1.03$ in the {entire computational domain}. Sensitivity analyses for grid, time-step, and domain sizes as well as a validation study of our computational methodology are provided in \cite{gungor-fish-swimming_wake}. More details of the presently utilized grid around the foils are presented in Figure \ref{fig_grid}. 

\begin{figure}
	\centering
	\includegraphics[width=1\textwidth]{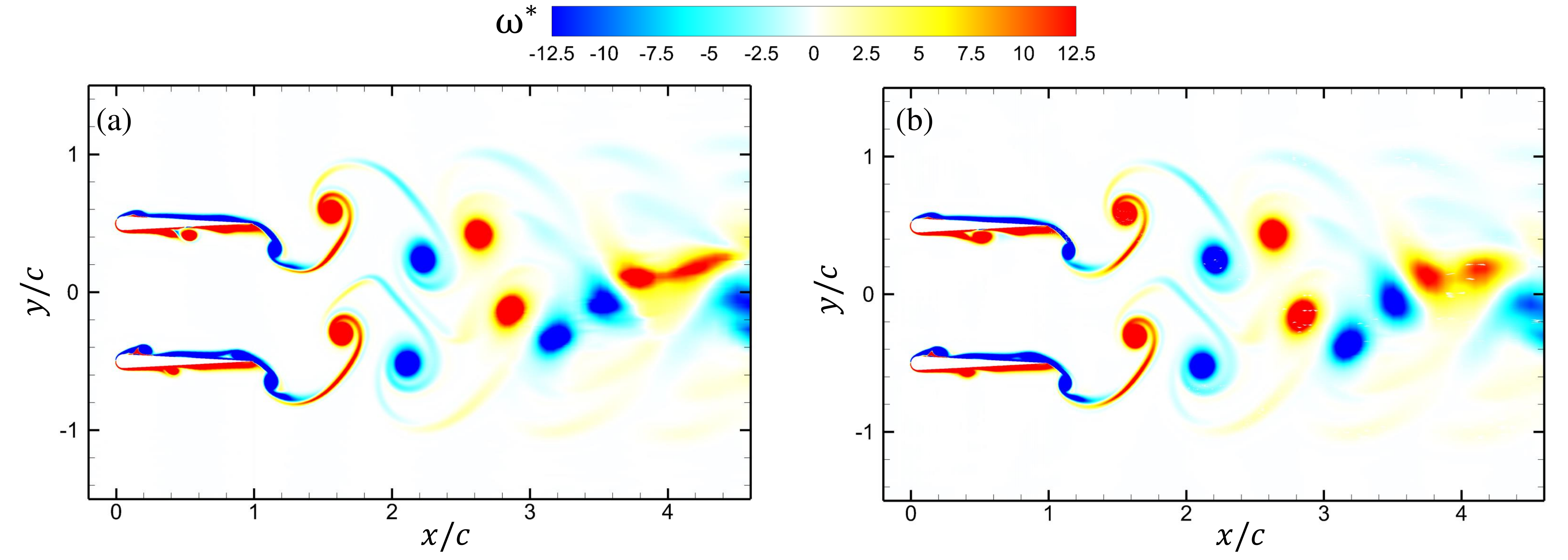}
	\caption{Comparing contour plots of spanwise component of vorticity ($\omega_z$) between (a) two- and (b) three-dimensional simulations for $Re=4000$ and $St=0.3$ at $t=10P$. The 3D case renders results on the center $xy-$plane.}
	\label{fig:2Dvs3D}
\end{figure}

Two-dimensional versus three-dimensional simulations are an important numerical complexity that can have implications on wake dynamics at high $Re$ flow conditions. To this effect, we carried out three-dimensional sensitivity studies to confirm that underlying physics of coherent structures in the flow, including wake deflection, wake merging, and vortex interactions, follow a two-dimensional or Q2D mechanism \citep{godoy-transitions-wake-2008, godoy-model-symmetry-breaking-2009, dewey-propulsive-performance-2014, shoele_synchornized_fins_2015, lagopoulos2020deflected}. \cite{deng_2d-3d_transition_2016} notes that 2D-to-3D transition in the wake of pure in-phase pitching foils occur at considerably high $St$, which excludes the parameter space employed here. Contour plots in Figure \ref{fig:2Dvs3D} compare coherent structures, and their interactions, along the center $xy-$plane of the wake at $Re=4000$, $St=0.3$, and $d^*=1$ with those from two-dimensional simulations. These results confirm that two- and three-dimensional simulations render very similar results in terms of coherent structures, wake dynamics and interactions related to merging. Moreover, it has been previously established that there are no significant variations observed between two- and three-dimensional cases in studying propulsive performance of pure pitching foils at moderate $St$, e.g., thrust, efficiency and power \citep{ zurman_3d_Effects_performance_2020}. This range comprises $St$ of the flow examined in the current study. Note that the impact of three-dimensionality on wake structures \citep{deng_2d-3d_transition_2016} and performance \citep{zurman_3d_Effects_performance_2020} becomes remarkable at relatively lower $St$ for pure heaving foils.

The cycle-averaged coefficients of thrust ($\widetilde{C_T}$) and power ($\widetilde{C_P}$) together with Froude efficiency ($\eta$) are calculated to discuss propulsive performance of the foils. These parameters are defined as:
\begin{equation}
\widetilde{C_T}=\frac{\widetilde{F_x}}{{\textstyle \frac{1}{2}} \rho U_\infty^2 sc},\label{eq:thrust}
\end{equation}	
\begin{equation}
\widetilde{C_P}=\frac{\widetilde{M_z}\dot{\theta}}{{\textstyle {1 \over 2}} \rho U_\infty^3 sc},\label{eq:power}
\end{equation}	
\begin{equation}
\eta=\frac{\widetilde{C_T}}{\widetilde{C_P}}.\label{eq:efficiency}
\end{equation}	
 Here, $\widetilde{F_x}$ is the streamwise force applied by the foil to the fluid, $\widetilde{M_z}$ is the moment in $z$ direction applied to the foil, $\rho$ is fluid density, $s$ is span of the foil. $\widetilde{F_x}$ and $\widetilde{M_z}$ are averaged within each oscillation cycles over at least 3500 timesteps.

\section{Results \& Discussion}
\label{ch_results}
We begin our analysis with examining vortex dynamics and wake interactions of parallel pitching foils. In a previous study \citep{gungor-asymmetry}, we examined transient wake developments of foils, performing in-phase and out-of-phase pitching in side-by-side configurations for $\mbox{St}=0.25-0.5$ and $d^*=1$ at $\mbox{Re}=4000$. It was demonstrated that wake structures showed quasi-steady characteristics and were in perfect agreement with the findings of \cite{dewey-propulsive-performance-2014}, i.e., merging symmetric wake for in-phase pitching and diverging symmetric wake for out-of-phase pitching foils. However, wake structures and propulsive performance of both in-phase and out-of-phase pitching foils were observed to be highly transient. The wake of in-phase pitching foils initially consisted of two deflected vortex streets parallel to each other. These streets merged after some time and formed a symmetric wake. The merging process coincides with the enhancement in  time-averaged thrust and efficiency of the foils. The opposite phenomena was observed in the wake of out-of-phase pitching foils. The foils initially produced diverging symmetric wakes whose symmetry was broken after several oscillation cycles. Here, we expand the parameter space to introduce and build classifications of the vortex patterns, which elucidate active flow control techniques possibly employed by natural swimmers, to gain a desired hydrodynamic performance. Furthermore, we also present a quantitative explanation for underlying physical mechanisms of the wake merging phenomenon.

\subsection{Classification of Vortex Patterns}

\begin{figure*}
	\centering
	\includegraphics[width=1\textwidth]{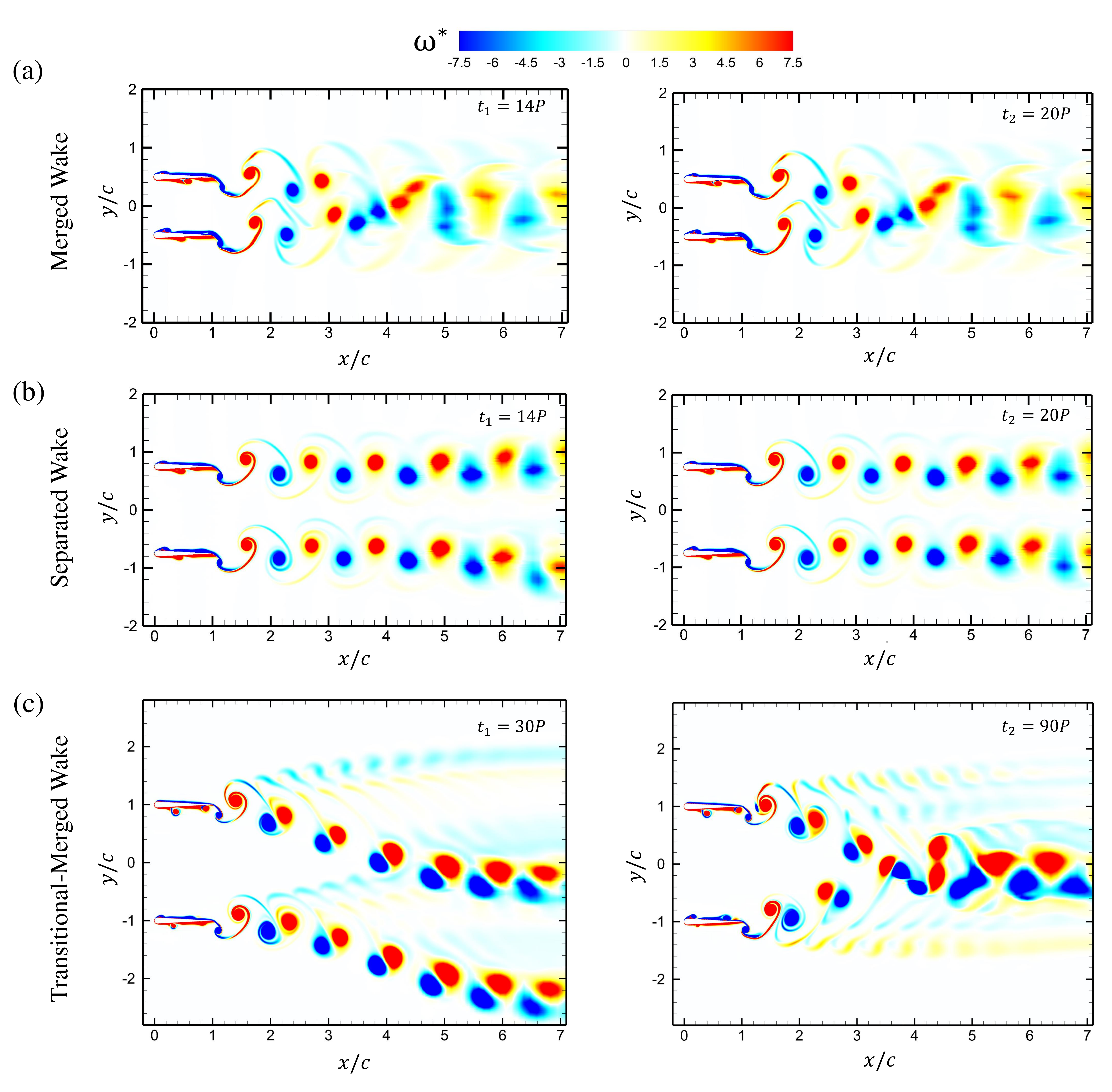}
	\caption{Contour of spanwise vorticity ($\omega_{z}^{*}$) of parallel foils for (a) $St=0.25$ and $d^*=1$ (merged wake), (b) $St=0.3$ and $d^*=1.5$ (separated wake), and (c) $St=0.5$ and $d^*=2$ (transitional-merged wake) at different time instants. Here, Vorticity is normalized by $U_{\infty}/c$.  (See supplementary videos: SV1, SV2, and SV3, for the entire wake evolution of (a), (b), and (c), respectively.)}
	\label{fig_classification_contours}
\end{figure*}

\begin{figure*}
	\centering
	\includegraphics[width=0.9\textwidth]{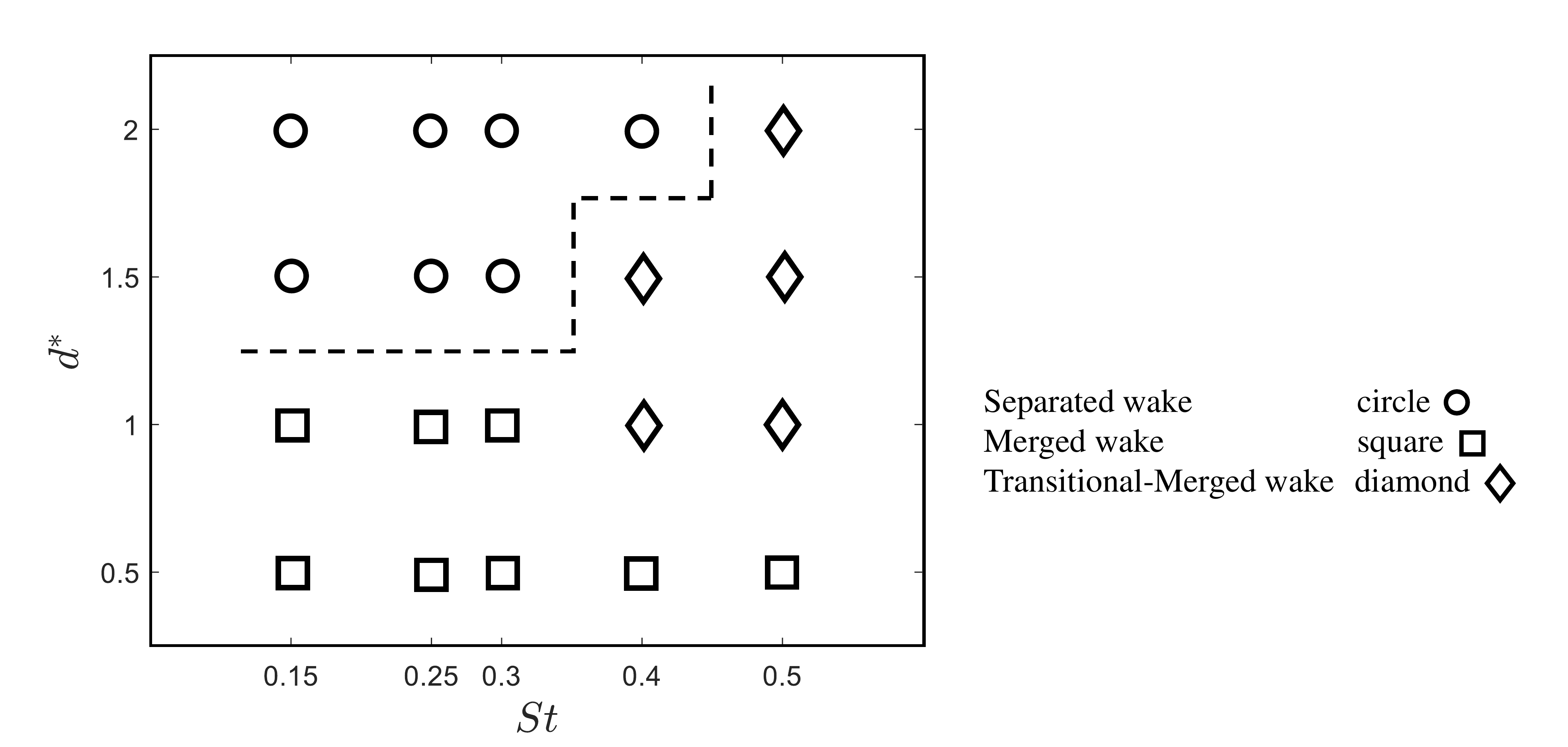}
	\caption{Classification of the wake patterns of in-phase pitching foils in side-by-side configuration at a range of separation distance and Strouhal number for $Re=4000$. Dashed lines correspond to the boundary that distinguished merged and separated wakes.}
	\label{fig_classification_diagram}
\end{figure*}

We identify three distinct vortex patterns in the wake of in-phase parallel pitching foils (side-by-side configuration) for the given parameter space, i.e., \mbox{$0.5 \geq d^* \geq 2$} and \mbox{$0.15 \geq St \geq 0.5$}. Merged$-$separated characteristics of the wakes were taken into consideration when classifying the wake in Figure \ref{fig_classification_contours}. Here, a merged wake corresponds to the vortex topology that involves the vortex streets shed by upper and lower foils merging in mid-wake and forming a single street, which constitutes a new flow configuration. In separated wakes, on the other hand, upper and lower vortex streets remain isolated without interacting with each other. Both merged and separated wakes demonstrate quasi-steady characteristics as the vortex patterns are formed within several pitching cycles and remain stable during the next oscillation cycles without altering significantly ({compare $t_1=14P$ and $t_2=20P$ of Figure \ref{fig_classification_contours}a and \ref{fig_classification_contours}b, where P is period of the pitching cycle).} Conversely, transitional-merged wakes undergo distinct separation and merging stages, transitioning from the former to the latter configuration. As explained earlier, oscillating foils produce deflected reverse BvK vortex streets at considerably high $\mbox{St}$. In the wake of parallel foils, interaction between  vortex streets shed by each foil results in the constitution of the symmetric wake, in which upper and lower wakes amalgamate around the centerline. (see Figure \ref{fig_classification_contours}c). The pitching cycle, in which the merging takes place, greatly depends on $d^*$ and $St$ (see Table \ref{table_merging_location}). For the sake of comparison, the merging process of the wakes occurs around 22\textsuperscript{nd} cycle for $d^*=1$ and $\mbox{St}=0.5$, whereas more than 75 cycles are needed for this phenomenon to occur for $d^*=2$ and $\mbox{St}=0.5$. These vortex patterns were gathered in a $St-d^*$ phase diagram in order to provide a thorough classification of the wakes of parallel foils in Figure \ref{fig_classification_diagram}. In this diagram, separated and merged wakes are observed in upper ($d^*\geq 1.5$) and lower ($d^*\leq 1$) regions of the diagram, respectively. On the other hand, transitional-merged wakes fall into the high $\mbox{St}$ region. It is important to note that this type of wakes are formed only at sufficiently high $\mbox{St}$ that facilitate the formation of deflected wakes. The relation between the deflection phenomena and wake merging will be further explained in the next part of this section.

\begin{figure*}
	\centering
	\includegraphics[width=1\textwidth]{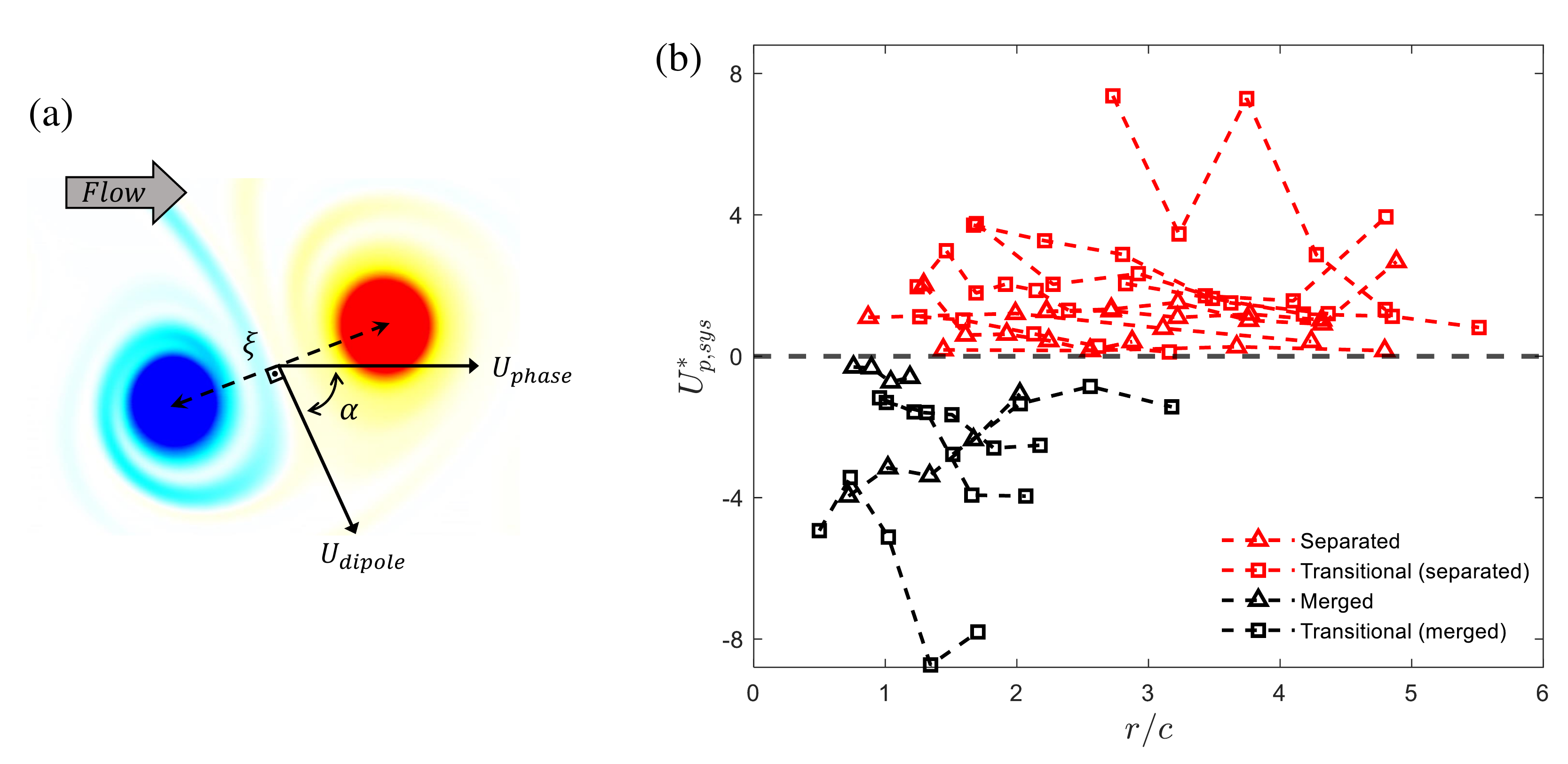}
	\caption{(a) Demonstration of the parameters used in the proposed model. (b) Effective phase velocity of the system with respect to displacement of the dipoles.}
	 \label{fig_classification_model}
\end{figure*}

Thereafter, we present quantification of important characteristics of the wakes and propose a model that disfperftinguishes emerging vortex patterns. At this point, it is important to recall the dipole model by \cite{godoy-model-symmetry-breaking-2009}, which presented a quantitative threshold for the wake deflection behind a single oscillating foil. This model also holds true for the wake of undulating foils \citep{khalid_undulating_deflection}. The wake of oscillating foils that consist of a reverse BvK vortex street is dominated by shedding of a counter-clockwise (positive sign) and a clockwise (negative sign) vortex per oscillation cycle that are located slightly above and below the centerline, respectively \citep{koochesfahani-vortical-patterns-1989}. The structure that consists of these two vortices is called dipole. Circulations of the vortices in the dipole induce velocity normal to the line that connects the vortex centers as described by the two-dimensional Bio-Savart rule \citep{Naguib_Vortex_Array_Model_2011}. When the self-advection velocity of the dipole is strong enough, it diverts the dipole from the centerline, which is followed by the consecutive dipoles. Therefore, the model was based on the offset between advection velocity of the propulsive wake, i.e., $U_{phase}$, and self-induced translation velocity of the dipole, i.e., $U_{dipole}$. They can be mathematically defined as follows:
\begin{equation}
U_{phase}=dX_i/dt,
\label{eq_U_phase}
\end{equation}
\begin{equation}
U_{dipole}=\Gamma / 2 \pi \xi,
\label{eq_U_dipole}
\end{equation}
where $X_i$ is the x-coordinate of {the spatial position of a} vortex core, $\Gamma$ denotes the circulation of counter-rotating vortices, and $\xi$ represents the distance between the centers of the vortices (see Figure \ref{fig_classification_model}a). Circulation ($\Gamma$) is computed either from a line integral of the velocity field or from a surface integral of voricity over the area bounded by a closed curve. \cite{godoy-model-symmetry-breaking-2009} used a rectangular frame, whose size was determined by Gaussian fit, to extract the boundary of each vortex towards calculating $\Gamma$. However, this method has a downside of potential numerical errors due to the possibility that rectangular frames may include counter-rotating vortices, particularly in the case of structures traversing in close proximity of one another. Therefore, we use an arbitrary closed curve to accurately capture each vortex proposed earlier by \cite{khalid_undulating_deflection}, which eliminates this error in computing $\Gamma$. The boundary of the curve is defined such that it only encompasses the region with magnitude of vorticty ($\omega$) greater than $5\%$ of its value in the flow field. Then, we determine the circulation of each vortex by calculating the line integral of the velocity field around the curve using the following definition:
\begin{equation}
\Gamma=\oint V \cdot dl.
\label{eq_circulation}
\end{equation}
For the effective phase velocity ($U^*_{p}$), \cite{godoy-model-symmetry-breaking-2009} defined it in the following manner that yields positive values for deflected wakes:
\begin{equation}
U^*_{p}=(U_{phase}-U_{\infty}) cos\alpha - U_{dipole},
\label{eq_U_effective}
\end{equation} 
\noindent where $\alpha$ is the angle between $U_{phase}$ and $U_{dipole}$ as presented in Figure \ref{fig_classification_model}a.

Here, we present a model that distinguishes different classes of vortex patterns using $U^*_{p}$. Although the model of \cite{godoy-model-symmetry-breaking-2009} successfully predicts whether the wake is deflected behind an isolated oscillating foil, but it cannot properly identify the nature of vortex patterns, i.e., merged or separated, for multiple parallel foils that exhibit wakes in close proximity of one another. In order to construct an effective mathematical model, our current work focuses on differentiating merged and separated wakes and supplying information about the direction of their deflections. To illustrate it further, transitional-merged wake at \mbox{$St=0.5$} and \mbox{$d^*=2$} exhibits deflection during each stage of wake development. During the separated stage at $t_1=30P$, both top and bottom wakes are deflected downwards, whereas the wake fully transitions to that of a merged configuration at $t_2=90P$. In the latter stage, upwards deflected bottom vortex street and downwards deflected top vortex street is observed (see Figure \ref{fig_classification_contours}c). However, the model proposed by \cite{godoy-model-symmetry-breaking-2009} cannot properly distinguish the vortex patterns for these configurations since all the cases consist of deflected wakes. Similarly, separated wakes at \mbox{$St=0.4$} and \mbox{$d^*=2$} (see Figure \ref{fig_merging_secondary_structures}a), which produce deflected vortex streets, and at lower \mbox{$St$} and $1.5 \leq d^* \leq 2$, which exhibit horizontal vortex streets (e.g. Figure \ref{fig_classification_contours}b), are treated disparately, while they are all classified as separated wake given the wake deflection is present only in a single case. Thus we introduce the term $\sin{\alpha}$ to the formulation to take the direction of deflection into account, because $\sin{\alpha}$ yields positive values for upwards wakes and negative values for downwards and non-deflected wakes. Thus, the \textit{effective phase velocity of the coupled vortex system} or $U^*_{p,sys}$ can now be defined as follows:
\begin{equation}
U^*_{p,sys}=\frac{U^*_{p,upper} \sin{\alpha}}{U^*_{p,lower} \sin{\alpha}}
\label{eq_U_effective_system}
\end{equation} 
Here, $U^*_{p,upper}$ and $U^*_{p,lower}$ are the effective phase velocities computed for upper and lower vortex streets. This mathematical relation produces negative $U^*_{p,sys}$ values for merging wakes due to the opposite deflection directions. In order to examine this model using our results, vortex dipoles are tracked as they move downstream the wake and $U^*_{p,sys}$ is calculated with respect to the distance dipoles travel after shedding and full separation from the foils. The displacement of these coherent structures is given by $r=\sqrt{(X_1-X_0)^2+(Y_1-Y_0)^2}$, where $X_1$ and $Y_1$ describe the instantaneous location of a vortex core. $X_0$ and $Y_0$ provide location of the vortex core just after its full detachment from the foils. Note that geometric mean of the respective quantities for the counter-rotating vortices is used as the location of the dipole. It is evident from Figure \ref{fig_classification_model}b that the proposed model works perfectly for the given parameter space. 

In this model, separated and merged stage of transitional-merged wakes are treated individually and marked with different colors, since these stages are contradictory to one another in terms of their vortex configuration. It is important to note that merged wake cases are tracked for relatively short radial displacement, i.e., \mbox{$r/c\leq 3$}. This is because their upper and lower vortex streets merge at mid-wake, which inhibits further tracking. However, dipoles of the separated wakes are traceable until circulation of the vortices shrink to negligible values due to the viscous diffusion around \mbox{$r/c=5$}. To clarify the working mechanism of the model, $U^*_{p}$ of merged wakes (see Figure \ref{fig_classification_contours}a or Figure \ref{fig_classification_contours}c at $t_2=90$) yield positive values for both bottom and top vortex streets as they are deflected upwards and downwards, respectively. Furthermore, $\sin{\alpha}$ for top and bottom wakes switch signs ($\sin{\alpha}<0$ for top and $\sin{\alpha}>0$ for bottom), which results in $U^*_{p,sys}<0$. On the other hand, horizontal vortex streets in the separated wakes (see Figure \ref{fig_classification_contours}b) have $U^*_{p}<0$ and $\sin{\alpha}<0$, thus leading to $U^*_{p,sys}>0$. Finally, the separated wake, whose vortex streets are deflected (see Figure \ref{fig_merging_secondary_structures}a), or transitional-merged wake at separated stage (see Figure \ref{fig_classification_contours}c at $t_1=30P$) yield $U^*_{p}>0$ and $\sin{\alpha}<0$, which translates to $U^*_{p,sys}>0$.

\subsection{Mechanism of Wake Merging}

 \begin{figure}
	\centering
	\includegraphics[width=0.95\textwidth]{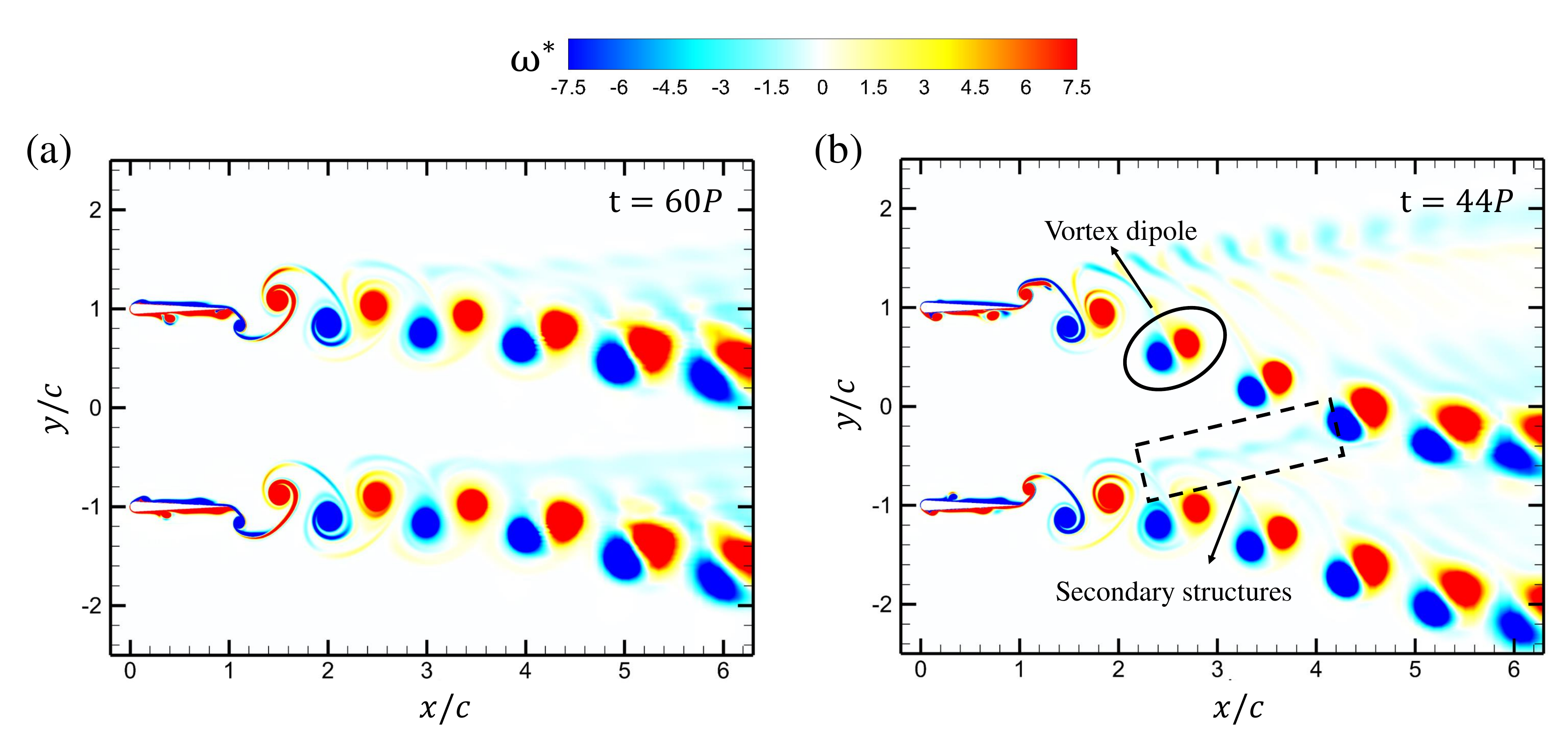}
	\caption{Contour of spanwise vorticity ($\omega_{z}^{*}$) of parallel foils at $d^*=2$ for (a) $St=0.4$ (separated wake) and (b) $St=0.5$ (transitional-merged wake) at different time instants. Here, vorticity is normalized by $U_{\infty}/c$. (See supplementary videos: SV3 and SV4 for the entire wake evolution of (b) and (a), respectively.)}
	\label{fig_merging_secondary_structures}
\end{figure}

Now that we have established a mathematical model to quantitatively identify and characterize different wake patterns behind parallel foils, we focus our attention to identify and explain the mechanism of wake merging. To this effect, we analyze the wake merging phenomenon by associating it with the production and dynamics of secondary vortex structures. When an oscillating single foil produces a deflected wake, secondary structures appear from the primary vortex street to move away from the direction of deflection \citep{gungor-fish-swimming_wake}. Such secondary structures were also observed in the experiments of \cite{godoy-transitions-wake-2008} and \cite{jones-knoller-betz-1998} and numerical simulations of \cite{liang-high-order-2011}. But no further analyses were conducted for this important feature of the wake dynamics. These structures are considerably weaker in their relative strength compared to those in the main street, which is why they have not received adequate attention in literature. We hypothesize that secondary structures play a key role in the merger of upper and lower vortex streets behind parallel oscillating foils. Figure \ref{fig_merging_secondary_structures}a shows separated wakes of the foils for the cases with $\mbox{St}=0.4$ and $d^*=2$ at $t=60P$. Even though secondary structures are present in the wake, structures from the lower wake are convected in the downstream direction before reaching the upper wake. At $St=0.5$ and $d^*=2$, in which wakes are merged, however, these secondary structures from the lower foil deflect upward to interact with primary street of the upper foil, traversing downward (see  Figure \ref{fig_merging_secondary_structures}b). These observations hint that this interaction triggers the wake merging process, because constructive or destructive interference of secondary vortices with the bigger coherent structures change their strengths in terms of circulation \citep{zhu2002three, akhtar2007hydrodynamics, khalid2021anguilliform, khalid2021larger}. Furthermore, onset of the complete wake merging appears located very close to the point of interaction between secondary structures and primary street. For instance, secondary structures shed by the lower foil at $St=0.5$ and $d^*=1.5$ catches the upper main vortex street at $x/c \approx 3$, which coincides with the location for the merging of wakes (please see Figure \ref{fig_classification_contours}c). Likewise, both the interactions between secondary structures and the upper wake as well as the merging of upper and lower wakes occur at $x/c \approx 4.65$ for $St=0.5$ and $d^*=2$ (please see the supplementary video and Table~\ref{table_merging_location}). 
The alignment of spatial merging location and that of vortex interactions strengthens our argument on the role of these smaller structures in initiating and facilitating the wake merging process.

\begin{figure*}
	\centering
	\includegraphics[width=0.475\textwidth]{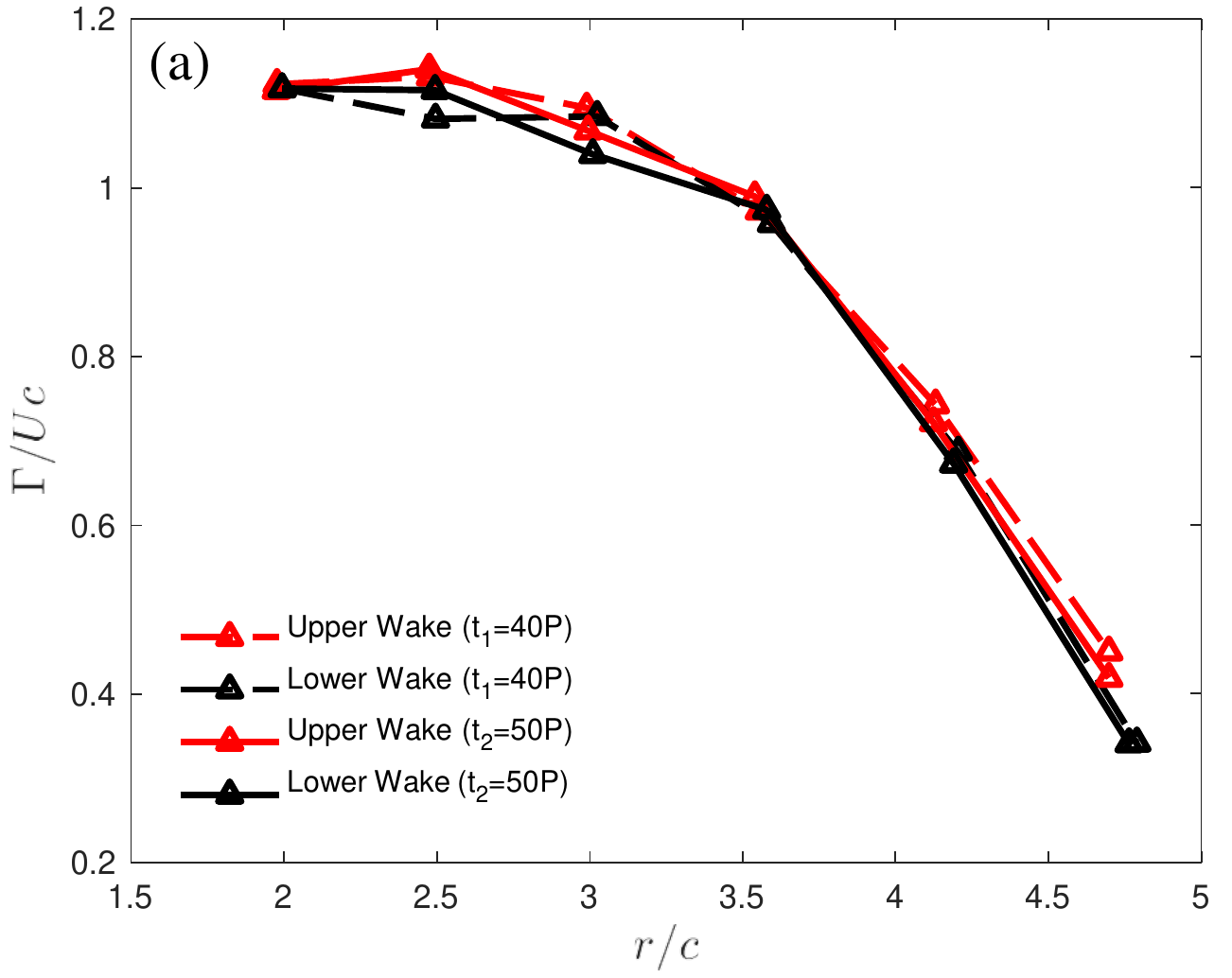}
	\hspace{0.2cm}
	\includegraphics[width=0.475\textwidth]{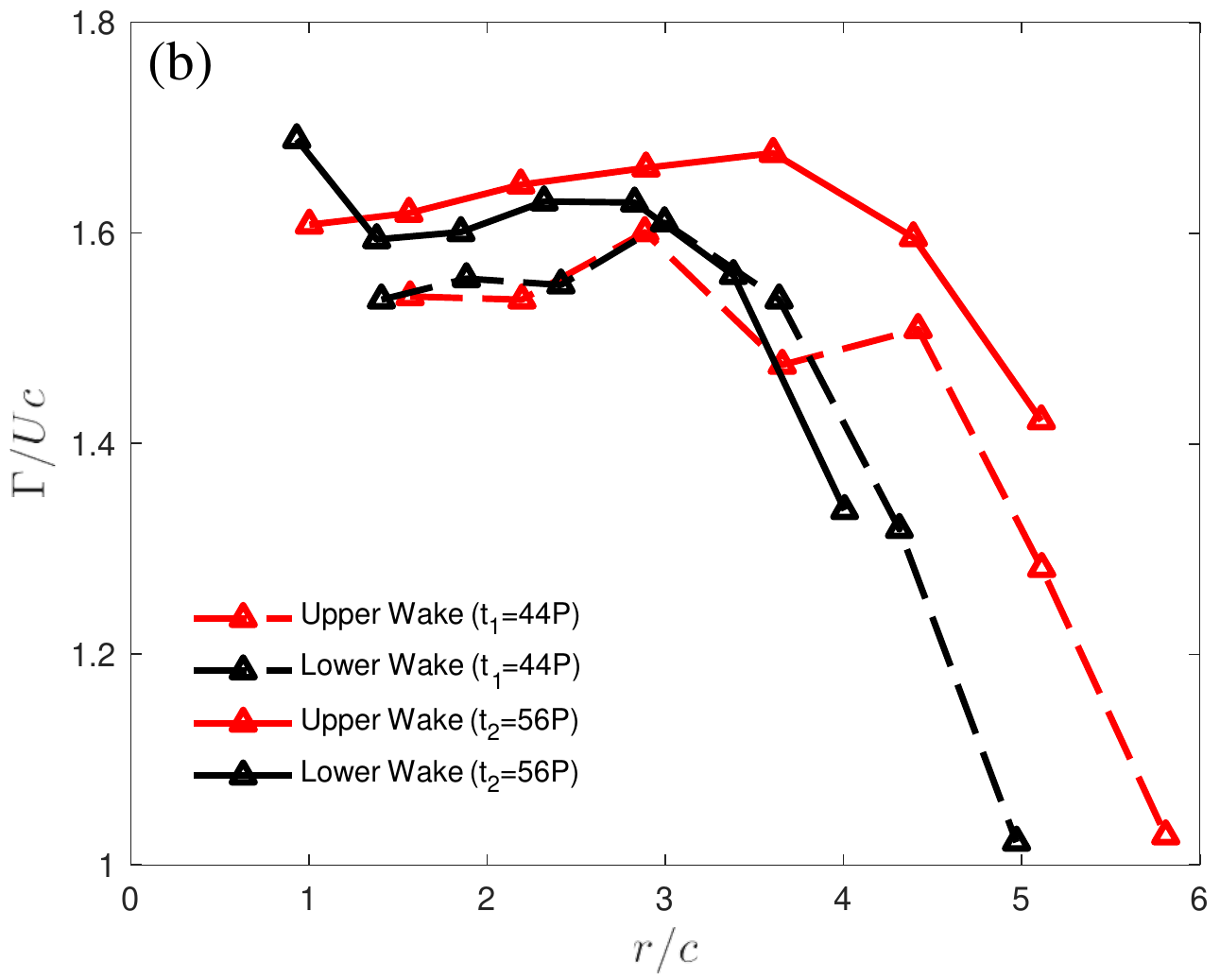}
	\caption{Magnitude of non-dimensional circulation of negative vorticity of upper and lower vortex streets for (a) $St=0.4$ (separated wake) and (b) $St=0.5$ (transitional-merged wake before the merger) at different time instants.}
	 \label{fig_merging_quantitative}
\end{figure*}

\begin{figure*}
	\centering
	\includegraphics[width=0.95\textwidth]{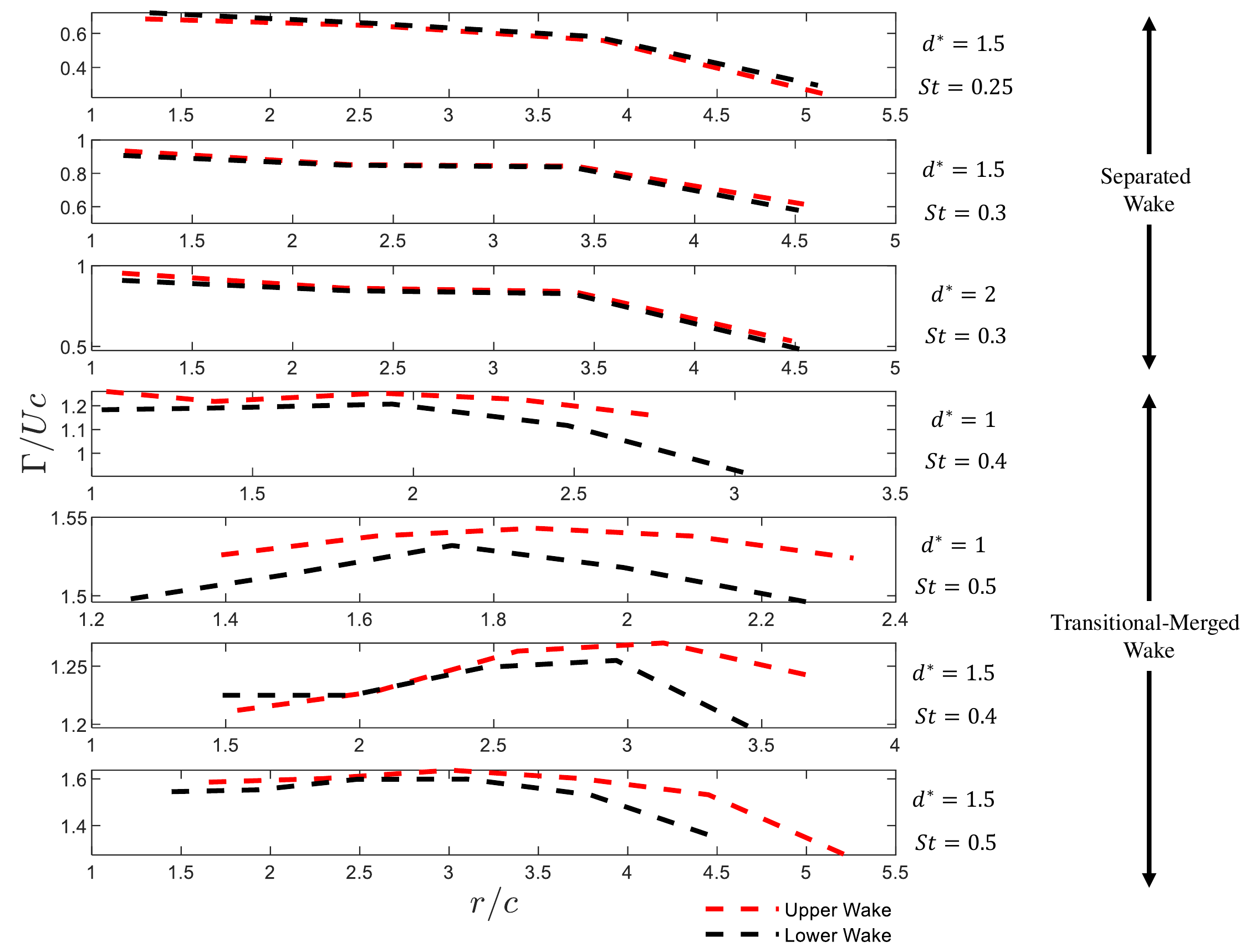}
	\caption{Magnitude of non-dimensional circulation of negative vorticity of upper and lower vortex streets for transitional-merged wakes before the merger and separated wakes.}
	 \label{fig_merging_quantitative_all_cases}
\end{figure*}

We proceed with providing another quantitative explanation to the impact of secondary structures on the overall wake dynamics. In this manner, locations and circulation of vortex dipoles are traced in the wake to provide evidence for the impact of secondary structures. Figure \ref{fig_merging_quantitative} exhibits the change of $\Gamma$ for negative vortices associated with separated ($St=0.4- d^*=2$) and transitional-merged ($St=0.5-d^*=2$) wakes, as dipoles move downstream the wake. For the transitional-merged wake (Figure \ref{fig_merging_quantitative}b), circulation of the negative vortices of upper and lower wakes overlap in the near wake region. However, proximity in this sense is broken in favor of the upper wake, after which there is constructive interference  of secondary structures with negative vorticity at $r/c \approx 3.5$. This observation remains valid at different time instants. Non-dimensional circulation at $t_1=44P$ and $t_2=55P$ is presented in Figure \ref{fig_merging_quantitative}b. On the contrary, there exists no significant difference in circulation of the upper and lower wakes for the separated wake (Figure \ref{fig_merging_quantitative}a), because the secondary structures have no influence on the upper wake. Similarly, wakes at $t_1=40P$ and $t_2=50P$ show that this trend is independent of the wake evolution and time. Furthermore, circulation of negative vortices of the upper and lower wakes is computed for other separated wakes and transitional merged wakes to support the set explanation for this mechanism. It is evident from Figure \ref{fig_merging_quantitative_all_cases} that our illustration stays valid for all transitional-merged wake cases investigated in this study. 

\subsection{Effect of Wake Merging on the Propulsive Performance of the System}
\label{cp_propulsive_performance}

\begin{figure*}
	\centering
	\includegraphics[width=1\textwidth]{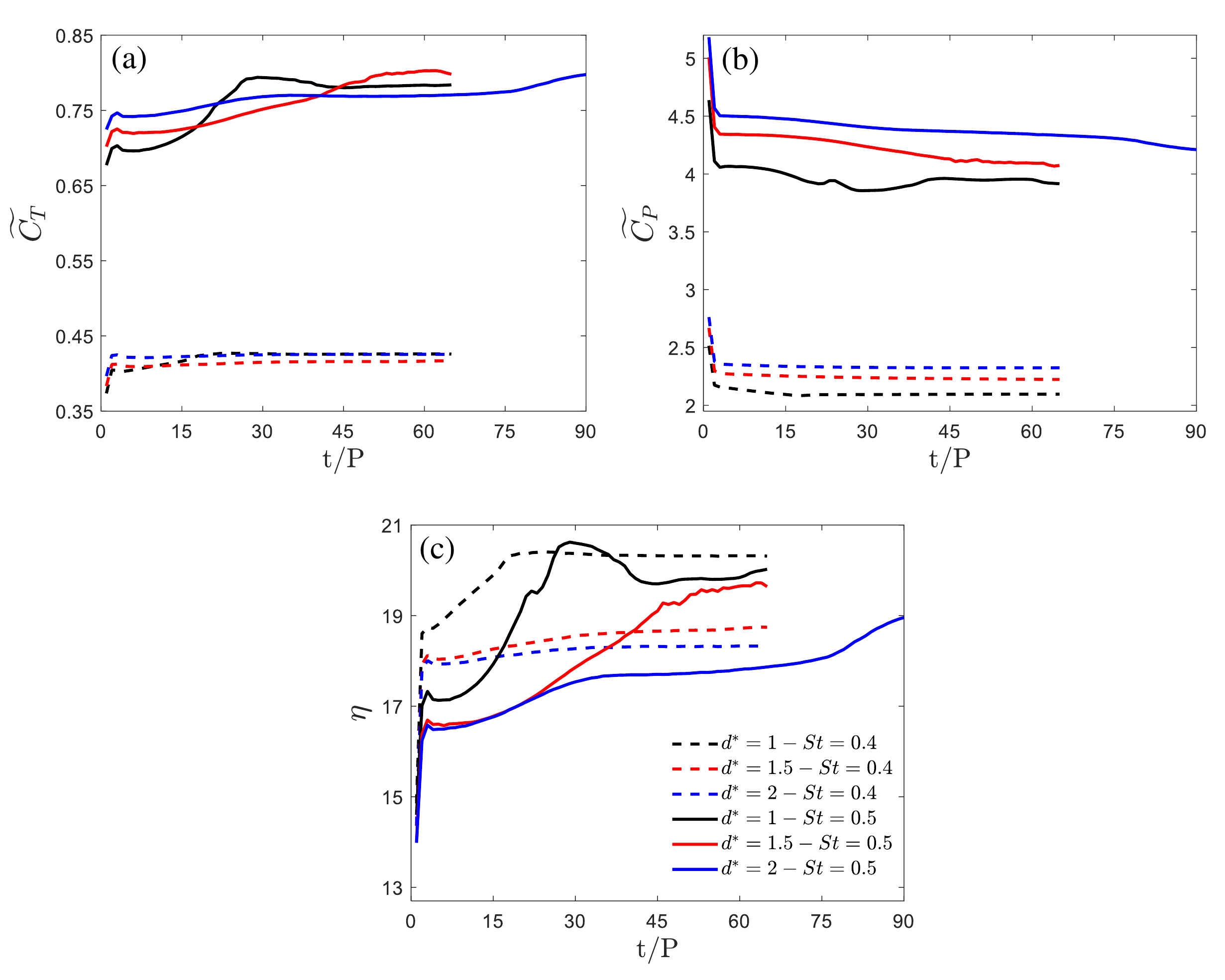}
	\caption{The variation of cycle-averaged (a) thrust and (b) power coefficients, as well as (c) efficiency of the system (averaged using Foil 1 and Foil 2) in time at a range of $St$ and $d^*$.}
	 \label{fig_merging_performance}
\end{figure*}

\begin{table*}
	\caption{\label{table_merging_location} Streamwise location ($x/c$) and time instant ($t/P$) in which the wake merging occurs as well as the percent improvement in the cycle-averaged coefficient of thrust ($\Delta \widetilde{C_T}$) for separated and transitional-merged wake cases at $St=0.4$ and $St=0.5$.}
	\centering
	\renewcommand{\arraystretch}{2.3}
	\setlength{\tabcolsep}{7.2pt}
\begin{tabular}{c|ccc|ccc|}
    &  \multicolumn{3}{c|}{$St=0.4$} & \multicolumn{3}{c|}{$St=0.5$} \\
    \hline
    & $d^*=1$ &  $d^*=1.5$ & $d^*=2$  & $d^*=1$ &$d^*=1.5$ &  $d^*=2$ \\
    \hline
    $x/c$  & 2.7 & 4.7 & $-$  & 2.5 & 3 & 3.6\\
	 $t/P$  & 15 & 49 & $-$  & 22 & 36 & 78\\
	 $\Delta \widetilde{C_T}$  & 5.15\% & 1.83\% & $0.95\%$  & 12.60\% & 10.77\% & 7.52\%\\
	\end{tabular}
\end{table*}

We now focus  on the relationship between propulsive performance of parallel foils (side-by-side configuration) and the wake merging phenomena by assessing the cycle-averaged performance metrics. Figure \ref{fig_merging_performance} shows temporal variations of the system-averaged (Foil 1 and Foil 2) coefficients of thrust and power, as well as efficiency at a range of $St$ and $d^*$. The performance parameters of this dynamical system is examined for $65$ oscillation cycles. The exception is the case of $St=0.5$ and $d^*=2$, in which the wake merging process occurs later at the $78^{th}$ cycle (see Table \ref{table_merging_location}). Thus, we examined this particular case using 90 cycles instead. Figure \ref{fig_merging_performance} covers all transitional-merged wakes observed in the current study as well as a separated wake, i.e. $St=0.4$ and $d^*=2$. In this section, we aim at establishing the impact of unsteady alterations in wake dynamics on the propulsive performance of the system. Thus, parameters for cases with lower \mbox{$St$}, i.e., \mbox{$St \leq 0.3$}, are not presented here as they display no significant wake transitions. Previously, it was demonstrated by \cite{gungor-asymmetry} that parallel foils generated highly quasi-steady performance and wake characteristics for lower \mbox{$St$} at $d^*=1$. With this background, It is further noticed in the present study that the same attributions persist for other separation distances examined here. Power requirements of the system marginally vary in time, which translates into resembling trends for the cycle-averaged coefficient of thrust and efficiency. The percent improvement in thrust calculated between $5^{th}$ and $65^{th}$ cycles ($5^{th}$ and $90^{th}$ for $St=0.5$ and $d^*=2$) is given in Table \ref{table_merging_location} together with the location and time instant of the wake merging. It is evident from Figure \ref{fig_merging_performance}a that generated thrust by the transitional-merged wakes improves with time and reaches a steady-steady after the wake is fully merged. This indicates that the wake merging process could be a contributing factor for thrust enhancement. Furthermore, it is important to note that the separated wake presented here ($St=0.4$ and $d^*=2$) experiences trivial alterations in propulsive performance parameters, which further supports our argument. Thus, we affirm that merging of these wakes improves propulsive thrust of the system by increasing thrust generation through amplification of the circulation associated with amalgamated vortices around the mid-wake. This results in the formation of high momentum jet on the centerline. There are two consequential inferences from Figure \ref{fig_merging_performance} and Table \ref{table_merging_location}, which strengthens our argument. First, the time instance of peaks in thrust variation lags the instant of wake merging process. For example, thrust generation for the case of $St=0.5$ and $d^*=1$ has its maximum at $t/P=29$, whereas its wake merging occurs at $t/P=22$. Second, thrust enhancement in the system decreases as the streamwise location of the onset of wake merging moves downtream. This is due to the reduction in circulation of the dipoles as they travel downstream the wake (see Figures \ref{fig_merging_quantitative} and \ref{fig_merging_quantitative_all_cases}). When upper and lower vortex streets merge in closer proximity to the foils, amalgamated vortices yield more circulation. This translates to an increased momentum jet that is induced by the vortex street. This is consistent with insignificant improvements in thrust for the case of $St=0.4$ and $d^*=1.5$, whose wake merging occurs farther in the wake, i.e., increased distance from the foils.

\begin{figure*}
	\centering
	\includegraphics[width=0.55\textwidth]{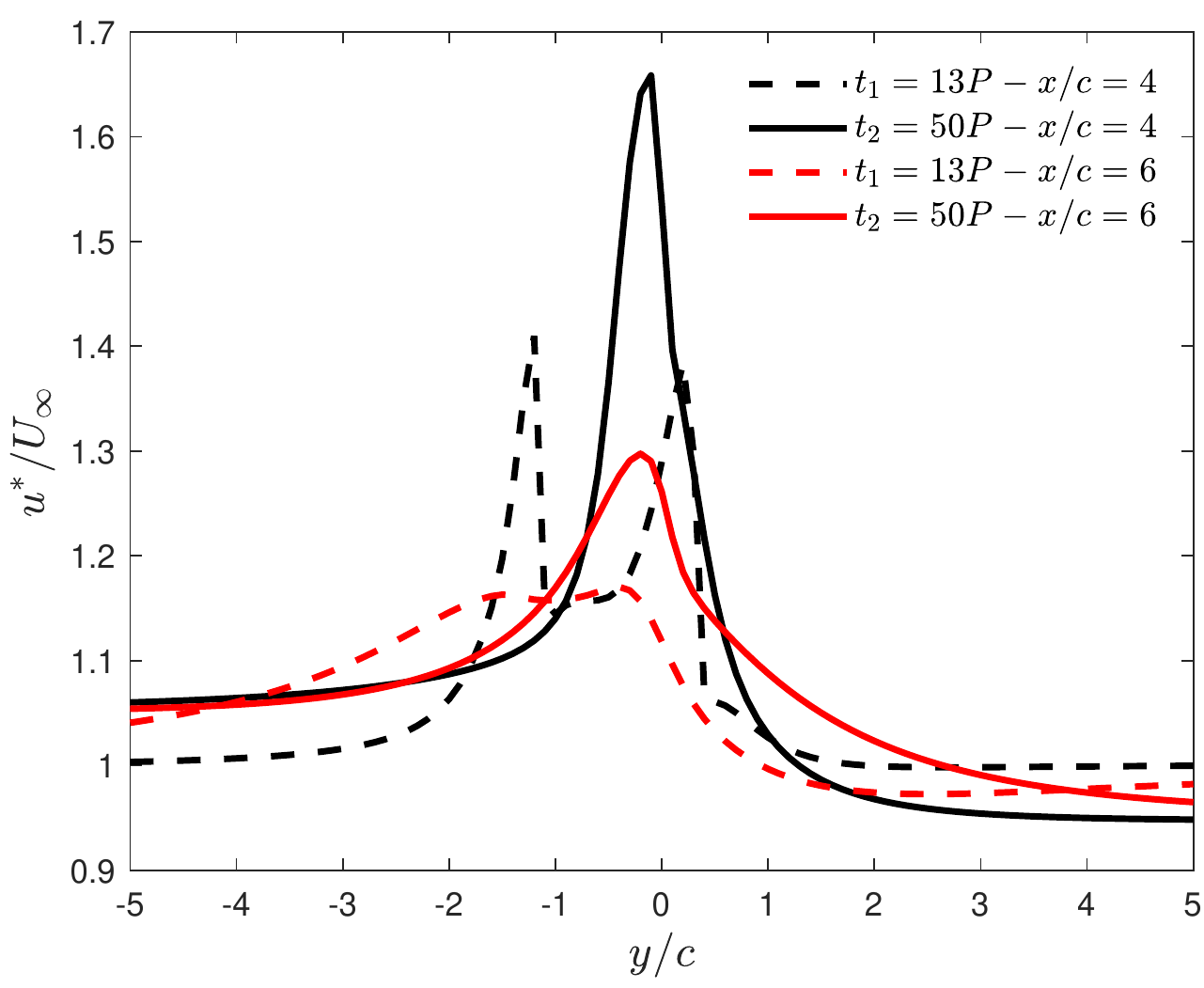}
	\caption{Cycle-averaged streamwise velocity ($u$) profiles, normalized by $U_{\infty}$, obtained from the finite-core vortex array model at $St=0.5$ and $d^*=1$ at different time instants ($t_1=13P$ and $t_2=50P$) and streamwise locations ($x/c=4$ and $x/c=6$).}
	 \label{fig_merging_u_velocity_profiles}
\end{figure*}

\begin{figure*}
	\centering
	\includegraphics[width=0.55\textwidth]{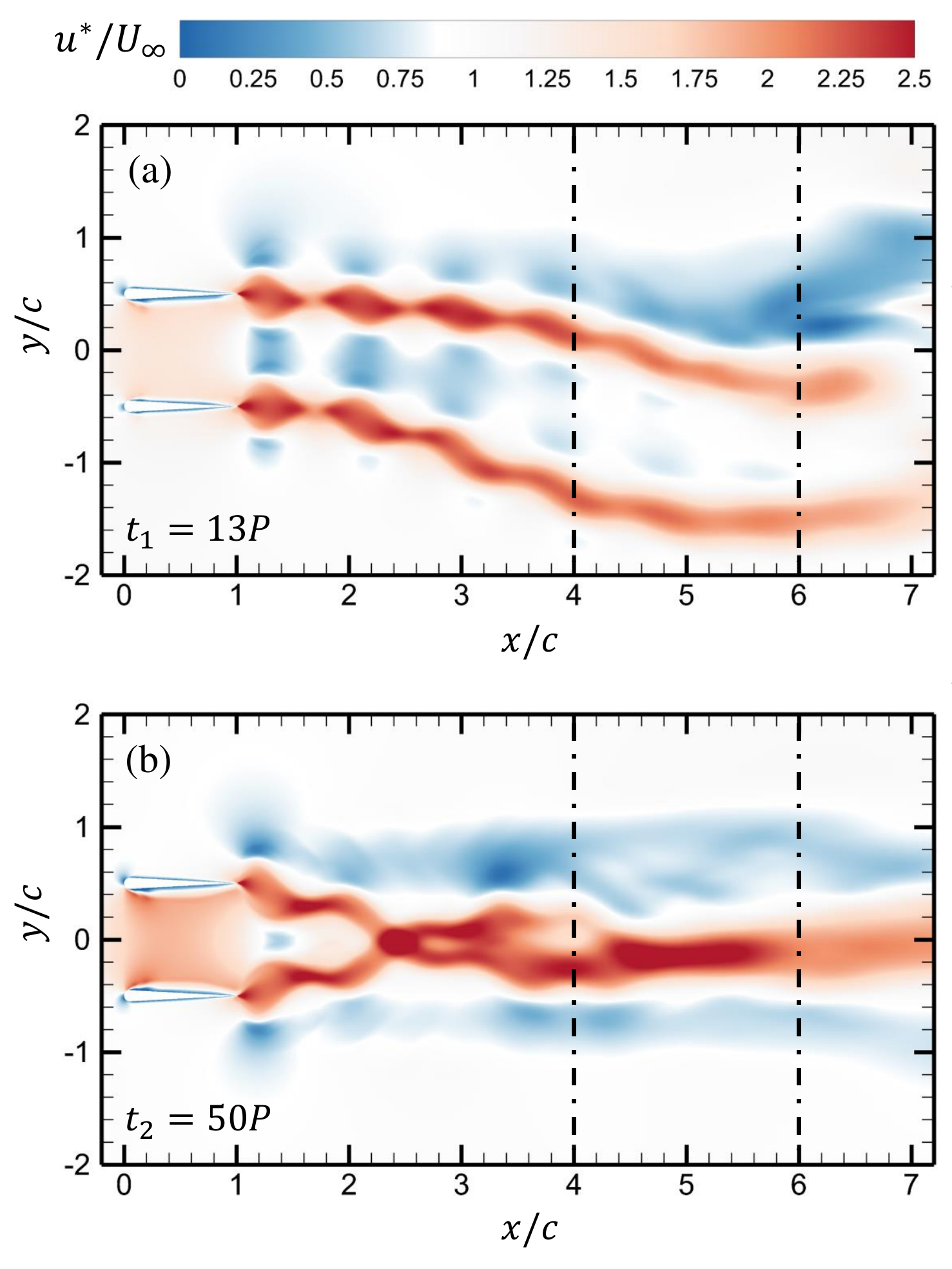}
	\caption{Cycle-averaged streamwise velocity ($u$) contours normalized by $U_{\infty}$ of in-phase pitching parallel foils at $St=0.5$ and $d^*=1$ for (a) $t_1=13P$ and (b) $t_2=50P$.}
	 \label{fig_merging_u_velocity_contours}
\end{figure*}

We now proceed with implementing the finite-core vortex array model to our wake data, following \cite{Naguib_Vortex_Array_Model_2011}. This model suggests that velocity profiles in the wake can be determined by superimposing finite amount of vortex cores onto a uniform flow. Streamwise and transverse velocity profiles that induced by superposition of $N$ vortices can be determined  using:
\begin{equation}
u(x,y)=U_{\infty}-\sum_{i=1}^{N}\frac{\Gamma_i (r_i)}{2\pi}\frac{(y-y_{ci})}{r_i^2},
\label{eq_vortex_array_u}
\end{equation}
\begin{equation}
v(x,y)=\sum_{i=1}^{N}\frac{\Gamma_i (r_i)}{2\pi}\frac{(x-x_{ci})}{r_i^2},
\label{eq_vortex_array_v}
\end{equation}
\\
where $r_i$ is the radial distance from $i^{th}$ vortex center to the point of calculation, $x_{ci}$ and $y_{ci}$ are the streamwise and transverse location of the center of the $i^th$  vortex, respectively. The model requires the number of vortices greater than or equal to 10 in order to converge \citep{Naguib_Vortex_Array_Model_2011}. To this end, locations and circulations of vortices within the range of $3 \leq x/c \leq 7$ are measured for $St=0.5$ and $d^*=1$ at $t_1=13P$ (separated stage) and at $t_2=50P$ (merged stage). The \mbox{$St$} and \mbox{$d^*$} of the flow are selected, considering it yields the highest percent improvement in thrust production (see Figure \ref{fig_merging_performance} and Table \ref{table_merging_location}). Moreover, the range of execution is determined considering that the wake merging occurs following $x/c=2.5$ and circulation of the dipoles diminish after $x/c \geq 7$. Mean streamwise velocity profiles calculated using the vortex array model is plotted for $x/c=4$ and $x/c=6$ in Figure \ref{fig_merging_u_velocity_profiles}. High velocity excess is observed around the centerline ($y/c=0$) for the merged wake ($t_1=50P$), whereas two distinct peaks associated with the jets created by the upper and lower vortex streets are visible in the separated wake ($t_1=13P$). Excess momentum at the outlet profiles ($x/c=4$ and $x/c=6$) created by the single velocity peak is greater than combination of the two peaks. It clearly shows impact of the wake merging on the formation of high-momentum jets, which translates to improvements in thrust generation. Furthermore, we compare velocity profiles obtained from the vortex array model with contours of mean horizontal velocity from Figure \ref{fig_merging_u_velocity_contours}. The model accurately captures the general trend and locations of the velocity peaks. However, it underestimates the magnitude of peaks in velocity profiles. This short-coming may be due to the sampling space given we only considered the region after wake merging in our calculations for the model. This focus on a particular region was driven by our emphasis on the influence of wake merging.

\begin{figure*}
	\centering
	\includegraphics[width=1\textwidth]{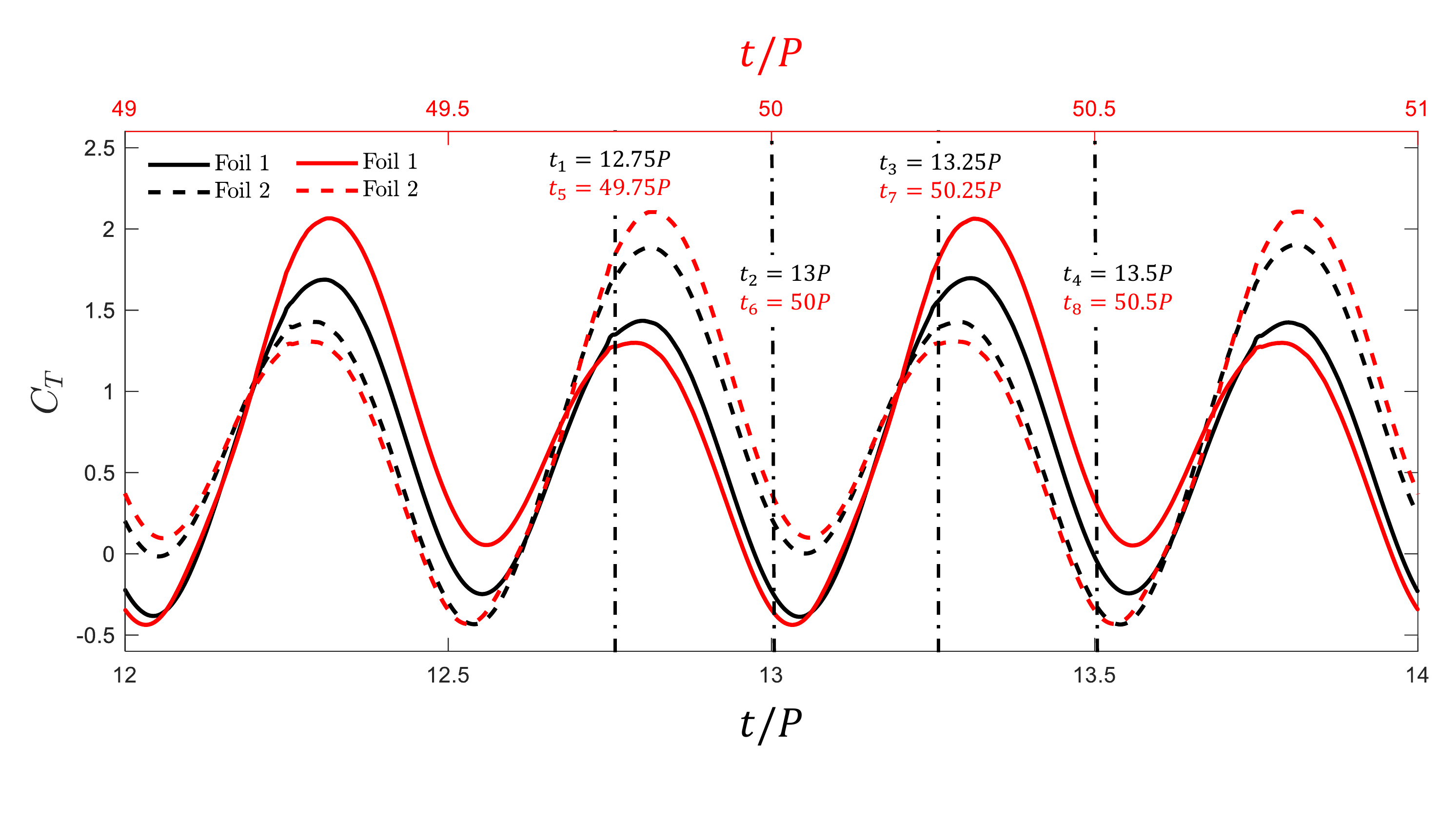}
	\caption{Variations in unsteady thrust coefficient of Foil 1 and Foil 2 for $St=0.5$ and $d^*=1$ (transitional-merged wake) during separated stage ($ 12 \leq t/P \leq 14$) illustrated in black and merged stage ($ 49 \leq t/P \leq 51$) of the wake evolution illustrated in red.}
	 \label{fig_merging_unsteady_CT}
\end{figure*}

\begin{figure*}
	\centering
	\includegraphics[width=1\textwidth]{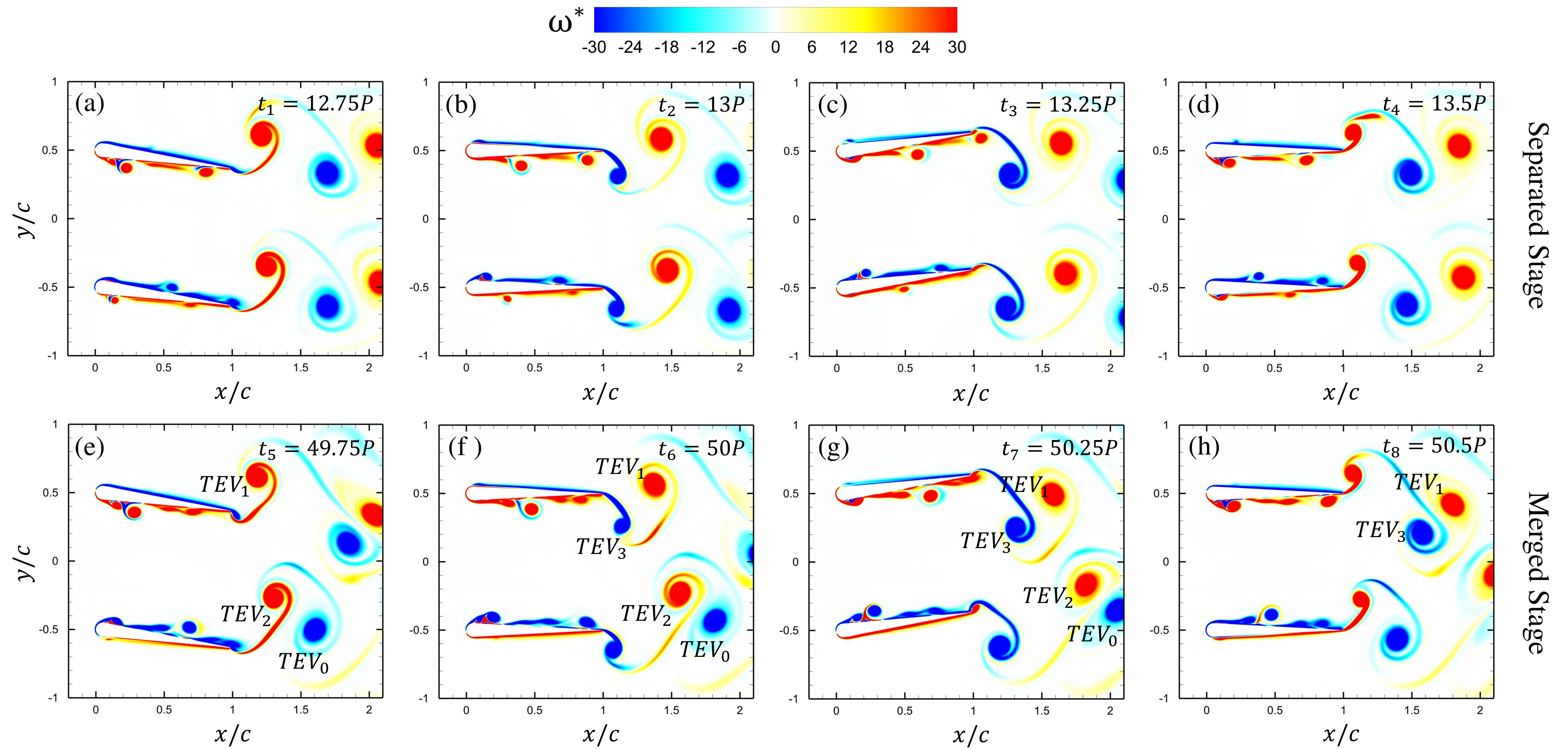}
	\caption{Contour of spanwise vorticity ($\omega_{z}^{*}$) around parallel foils for $St=0.5$ and $d^*=1$ (transitional-merged wake) at various time instants during separated stage: (a) $t_1=12.5P$, (b) $t_2=13P$, and merged stage: (c) $t_3=49.5P$, (d) $t_4=50P$. Here, Vorticity is normalized by $U_{\infty}/c$.  (See supplementary video: SV5 for the entire wake evolution)}
	 \label{fig_merging_LEV}
\end{figure*}

The formation, growth and interaction of LEVs can be an important mechanism that impact propulsive performance of oscillating foils \citep{anderson1998oscillating, pan_BEM_LEV_2012}, while also impacting the wake development \citep{hemmati-trailing-edge-shape-2019}. To this effect, we now look at how alterations in the formation and growth of LEVs around the foils influence their propulsive performance during separated-to-merged wake transition. LEVs are formed when the angle of attack is high enough that a separation bubble is formed on the foils. Their presence and evolvement on the surface of a fin or wing is responsible for a large part of thrust and lift generation in aquatic locomotion \citep{borazjani_LEV_2013, Bottom_LEV_stringrays_2016, liu_body-fin_2017, xiong_LEV_fin_2019} and insect flight \citep{Ellington_LEV_insect_1996, Birch_LEV_attachment_2001, Bomphrey_LEV_hawkmoth_2005}, respectively.  Unsteady thrust variations of the foils in the transitional-merged wake ($St=0.5$ and $d^*=1$) throughout the separated and merged time ranges is presented in Figure \ref{fig_merging_unsteady_CT}, which clearly demonstrates that higher peaks and lower troughs are achieved after the wake merging. Contours of vorticity focusing on surfaces of the foils at time instants that correspond to the highest ($\theta=8^{\circ}$) and the lowest ($\theta=0^{\circ}$) angles of attack is shown in  Figure \ref{fig_merging_LEV}. This enables comparing this process to the evolution of LEVs around the foils. Note that these instants roughly overlap with the times at which the foils yield their highest and lowest thrust generation, as marked in Figure \ref{fig_merging_unsteady_CT}. Negative (clockwise rotating) and positive (counter clockwise rotating) LEV are formed on the upper and lower surface of the foils at the separated stage of the wake evolution, respectively. On the other hand, constituting positive LEVs on the lower surface of Foil 1 and negative LEVs on the upper surface of Foil 2 is significantly suppressed when the wake is fully merged. Furthermore,  it is evident from contour plots  in Figure \ref{fig_merging_LEV} that stronger LEVs are formed after the wake merging, which hints at a potential factor for thrust enhancement. These observations are valid for the other transitional-merged wake cases with recognizable thrust enhancement as well, although they are not shown here for brevity. Note that profiles of unsteady thrust have two peaks and two troughs per oscillation cycle.  The impact of LEVs on thrust generation can be explained through the low pressure region (suction) formed by vortices. LEVs attached close to the anterior part of the foil favorably affect thrust by dropping the pressure in this region, whereas their influence is adverse if located around the posterior part of the foil. For example, Foil 1 at $t_7=50.25P$ have enhanced LEV around the front edge comparing to Foil 1 at $t_3=13.25P$. However, the distribution of LEVs on rear surfaces of the foils are matching, which translates to thrust enhancements for Foil 1. Likewise, thrust of Foil 2 at $t_5=49.75P$ is considerably larger than that of Foil 1. This is due to stronger LEVs formed close to the forehead of Foil 2, while an LEV  with large negative vorticity is present on the rear part of the Foil 1. 

It has been previously demonstrated by \cite{gungor-asymmetry} that merging of vortex streets is accompanied by the restoration of wake symmetry for parallel foils. Both performance and wake characteristics exhibit symmetric behavior with a delay of half pitching period. This lag between the foils is due to the formation process of vortex dipoles. Although TEV$_1$ and TEV$_2$ are shed at the same time (e.g., $t_5=49.75P$), TEV$_2$ establishes a dipole with TEV$_0$ that has been shed half a period prior to these TEVs, whereas the coupling of TEV$_1$ and TEV$_3$ are delayed by half a cycle. Therefore, distribution of LEVs around Foil 1 and Foil 2 at $t_5=49.75$ in Figures \ref{fig_merging_LEV}e is mirror image symmetric with switched signs of vorticity with that around Foil 2 and Foil 1 at $t_7=50.25P$ in Figures \ref{fig_merging_LEV}g, respectively. Besides, shifting $C_T$ values for Foil 1 by a half period in either direction results in a perfect overlap with those for Foil 2 during the merged stage of the wake evolution, or vice versa (see Figure \ref{fig_merging_unsteady_CT}). Contrarily, there is neither a coherent similarity of the arrangement of LEVs around the foils between different time steps nor a half period lag between performance parameters of Foil 1 and Foil 2 during the separated stage of wake evolution.

\section{Conclusions}
\label{cp_conclusion}

Numerical simulations on the flow around two in-phase pitching foils in side-by-side configuration are examined at a range of Strouhal number, $0.15<St<0.5$ and separation distance, $0.5<d^*<2$, at Reynolds number of 4000. First, we classified the vortex patterns in the wake. Separated and merged wakes, which exhibit quasi-steady performance and wake characteristics, were observed at lower Strouhal numbers. Small spacing between the foils yielded the constitution of merged wakes, while separated wakes are seen at higher separation distances. On the other hand, transitional-merged wakes, which are often observed at high Strouhal numbers, exhibited wake evolution in time. Two distinct and deflected vortex streets shed by each foil were observed at early stages of the oscillations. Upper and lower vortex streets approached each other in time, which eventually resulted in merging of the wakes. A novel mathematical model was proposed, which quantitatively established the threshold for the two set vortex patterns. This model utilized the locations and circulations of individual vortices in a dipole. It was further tested using the current parameter space and performed perfectly in determining if the wake is separated or merged. Then, we proceeded with evaluating and explaining the physical mechanism associated with the wake merging phenomenon. This analysis revealed a novel process in which secondary structures in the wake were responsible in part for the wake merging. The wake merging occurred when secondary structures from the lower vortex street were strong enough to form a constructive interaction with main vortex street of the upper wake. This interaction triggered the merging of wakes by increasing the circulation of negative vortex in the upper vortex street. In turn, this impacted the resultant induced velocity (flow) by the two vortex streets, which now do not match, leading to further deflection of wakes and their subsequent merger. Finally, it was observed that merging of the wakes enhances propulsive performance of the foils by combining circulations of amalgamated vortices. This process induces high-momentum jet around the centerline. Evolution of leading-edge vortices played a major role in the performance enhancement. Alterations in the distribution of leading-edge structures and the amplification in their strength, which occurred after the wake merging, was a contributing factor for the improvements in thrust generation.

\section*{Acknowledgments}

This research has received support from the Canada First Research Excellence Grant. The computational analysis was completed using Compute Canada clusters. 

\section*{Declaration of Interests}

The authors report no conflict of interests.

\bibliographystyle{jfm}
\bibliography{references}

\end{document}